\documentclass[a4paper,twoside,notitlepage]{article}
\usepackage[applemac]{inputenc}
\usepackage{natbib,epsf,amssymb,bm}
\usepackage{amsmath}
\usepackage{epsfig}
\usepackage{epstopdf}
\usepackage{SectsV9}% requires SectsV9.sty
\usepackage{float}

\newcommand{\bcen}{\begin{center}}
\newcommand{\ecen}{\end{center}}

\newcommand{\oxfont}{\fontfamily{cmr}\fontshape{n}
  \fontseries{m}\fontsize{9}{10}\selectfont}
\newcommand{\absfont}{\fontfamily{cmr}\fontshape{n}
  \fontseries{m}\fontsize{8}{9}\selectfont}
\newcommand{\autfont}{\fontfamily{cmr}\fontshape{it}
  \fontseries{m}\fontsize{11}{12}\selectfont}
\newcommand{\titfont}{\fontfamily{cmss}\fontshape{n}
  \fontseries{bx}\fontsize{17}{18}\selectfont}

\newcommand{\aut}[1]{\vspace*{2mm}\centerline{{\autfont\scshape #1}}}
\newcommand{\loc}[1]{\vspace*{2pt}\centerline{{\itshape #1}}}
\newcommand{\email}[1]{\vspace*{2pt}\centerline{{\ttfamily #1}}}

\newcommand{\running}[2]{\markboth{\hfill {{\itshape #1}}}
 {{\itshape #2}}\hfill}

\newcommand{\info}[1]{
 \renewcommand{\thefootnote}{\fnsymbol{footnote}}
 \footnotetext[0]{\kern-1.6\parindent #1}
 \renewcommand{\thefootnote}{\arabic{footnote}}}
\newcommand{\bg}{\;\bigg\vert\;}

\newenvironment{tit}{\bcen\titfont}{\ecen}
\newcommand{\btit}{\vspace*{18mm}\begin{tit}}
\newcommand{\etit}{\end{tit}}

\newenvironment{abs}
{\begin{quote}\absfont \bcen{\scshape Summary}\ecen \vspace*{1mm} }
{\end{quote}}
\newcommand{\babs}{\begin{abs}}
\newcommand{\eabs}{\end{abs}}
\newcommand{\ti}[1]{\emph{#1}}
\newenvironment{key}
{\begin{quote}\begin{flushleft}\absfont \ti{Keywords and Phrases:} 
  \scshape}{\end{flushleft}\end{quote}}
\newcommand{\bkey}{\begin{key}}
\newcommand{\ekey}{\end{key}\vspace*{1mm}}

\newcommand{\beqn}{\begin{eqnarray*}}
\newcommand{\eeqn}{\end{eqnarray*}}
\newcommand{\beqnn}{\begin{eqnarray}}
\newcommand{\eeqnn}{\end{eqnarray}}

\newenvironment{parag}{\par}{\par}
\newenvironment{dif}
  {\begin{parag}\absfont  \begin{parag}}
  {\end{parag}\end{parag}}
\newcommand{\bdif}{\begin{dif}}
\newcommand{\edif}{\end{dif}}

\newtheorem{teo}{Theorem}
\newcommand{\bteo}{\begin{teo}\itshape}
\newcommand{\bteon}[1]{\begin{teo}{\fsize{9}\mbf (#1).}\itshape}
\newcommand{\eteo}{\end{teo}}
\newcommand{\ok}{\hfill\raise-3pt\hbox{$\square$}\vspace*{1mm}}
\newenvironment{proof}
      {\begin{dif} \noindent{\em Proof.~}}
      {\ok\vspace*{15pt}\end{dif}}
\newcommand{\bpro}{\begin{proof}}
\newcommand{\epro}{\end{proof}}

\newcommand{\bsnuevea}{\thispagestyle{empty}\oxfont
        \vspace*{-\headsep}\vspace*{-\headheight}\vspace*{-10pt}
        \begingroup\absfont \itshape
        \noindent BAYESIAN STATISTICS 9, 
        \newline J.~M.~Bernardo, M.~J.~Bayarri, J.~O.~Berger,
        A.~P.~Dawid,\newline D.~Heckerman, A.~F.~M.~Smith
        and M.~West (Eds.)\newline
        \copyright\ Oxford University Press, 2010
        \endgroup}

\makeatletter

\floatstyle{ruled}
\newfloat{algorithm}{tbp}{loa}
\floatname{algorithm}{Algorithm}

\newcommand{\I}{\mathrm{I}}

\begin{document}

\bsnuevea

\btit Free energy Sequential Monte Carlo, application to mixture
modelling \etit 

\running{N. Chopin \& P. Jacob}{Free energy SMC}
\aut{N. Chopin \& P. Jacob}
\loc{CREST (ENSAE), France}
\email{nicolas.chopin@ensae.fr,\quad pierre.jacob@ensae.fr}
\info{Support from the ANR grant ANR-008-BLAN-0218 by the French
  Ministry of research is acknowledged.}

\babs
We introduce a new class of Sequential Monte Carlo (SMC) methods,
which we call free energy SMC. This class is inspired by free energy
methods, which originate from Physics, and where one samples from
a biased distribution such that a given function $\xi(\theta)$ of
the state $\theta$ is forced to be uniformly distributed over a given
interval. From an initial sequence of distributions $(\pi_t)$ of
interest, and a particular choice of $\xi(\theta)$, a free energy SMC sampler
computes sequentially a sequence of biased distributions $(\tilde{\pi}_{t})$
with  the following properties: (a) the marginal distribution
of $\xi(\theta)$ with respect to $\tilde{\pi}_{t}$ is approximatively uniform
over a specified interval, and (b) $\tilde{\pi}_{t}$ and $\pi_{t}$
have the same conditional distribution with respect to $\xi$. We
apply our methodology to mixture posterior distributions, which are
highly multimodal. In the mixture context, forcing certain hyper-parameters
to higher values greatly faciliates mode swapping, and makes it possible
to recover a symetric output. We illustrate our approach with univariate
and bivariate Gaussian mixtures and two real-world datasets. 
\eabs

\bkey
Free energy biasing; Label switching; Mixture; Sequential
Monte Carlo; particle filter.
\ekey

\section{Introduction}

A Sequential Monte Carlo (SMC) algorithm (a.k.a. particle filter)
samples iteratively a sequence of probability distributions
$(\pi_{t})_{t=\text{0,...,T}}$, through importance sampling and
resampling steps. The initial motivation of SMC was the sequential
analysis of dynamic state space models, where $\pi_{t}$ stands for the
filtering distribution of state (latent variable) $x_{t}$, conditional
on the data $y_{1:t}$ collected up to time $t$; see e.g. the book of
\cite{DouFreiGor}. Recent research however
\citep{Neal:AIS,Chopin:IBIS,DelDouJas:SMC} have extended SMC to
{}``static'' problems, which involves a single, but {}``difficult''
(in some sense we detail below) distribution $\pi$. Such extensions
use an artificial sequence $(\pi_{t})_{t=\text{0,...,T}}$, starting at
some {}``simple'' distribution $\pi_{0}$, and evolving smoothly
towards $\pi_{T}=\pi$. Two instances of such strategies are i)
annealing (\citealt{Neal:AIS}, see also
\citealt{gelman1998simulating}), where
$\pi_{t}(\theta)=\pi_{0}(\theta)^{1-\gamma_{t}}\pi(\theta)^{\gamma_{t}}$,
and $\gamma_{t}=t/T$, or some other increasing sequence that starts at
$\text{0}$ and ends at $1$; and ii) IBIS \citep{Chopin:IBIS}, where
$\pi$ stands for some Bayesian posterior density
$\pi(\theta)=p(\theta|y_{1:T})$, conditional on some complete dataset
$y_{1:T}$, and $\pi_{t}(\theta)=p(\theta|y_{1:t})$.  For a general
formalism for SMC, see \citet{DelDouJas:SMC}.

One typical ``difficulty'' with distributions of interest $\pi$
is multimodality. A vanilla sampler typically converges to a single
modal region, and fails to detect other modes, which may be of higher
density. The two SMC strategies mentioned above alleviate this problem
to some extent. In both cases, $\pi_{0}$ is usually unimodal and
has a large support, so ``particles'' (sampled points) explore
the sample space freely during the first iterations. However, this
initial exploration is not always sufficient to prevent the sample
to degenerate to a single modal region. We give an illustration of
this point in this paper. 

To overcome multimodality, the molecular dynamics community has
developed in recent years an interesting class of methods, based on
the concept of free energy biasing; see for instance the book of
\cite{TonyGab:book} for a general introduction. Such methods assume
the knowledge of a low-dimensional function $\xi(\theta)$, termed as
the {}``reaction coordinate'', such that, conditional on
$\xi(\theta)=x$, the multimodality (a.k.a.  mestability in the physics
literature) of $\pi$ is much less severe, at least for certain values
of $x$. The principle is then to sample from $\tilde{\pi},$ a free
energy biased version of $\pi$,
$\tilde{\pi}(\theta)=\pi(\theta)\exp\left\{ A\circ\xi(\theta)\right\}
$, where $A$ denotes the free energy, that is, minus log the marginal
density of the random variable $\xi(\theta)$, with respect to $\pi$.
This forces an uniform exploration of the random variable
$\xi(\theta)$, within given bounds. At a final stage, one may perform
importance sampling from $\tilde{\pi}$ to $\pi$ to recover the true
distribution $\pi$.

The main difficulty in free energy biasing methods is to estimate
the free energy $A$. A typical approach is to compute sequentially
an estimate $A_{(t)}$ of $A$, using some form of Adaptive MCMC (Markov
chain Monte Carlo): at each iteration $t$, a MCMC step is performed,
which leaves invariant $\pi_{(t)}(\theta)=\pi(\theta)\exp\left\{ A_{(t)}\circ\xi(\theta)\right\} $,
then a new estimate $A_{(t+1)}$ of the free energy is computed from
the simulated process up to time $t$. The simulation is stopped when
the estimate $A_{(t)}$ stabilises in some sense. Convergence of Adaptive
MCMC samplers is a delicate subject: trying to learn too quickly from
the past may prevent convergence for instance. These considerations
are outside the scope of this paper, and we refer the interested reader
to the review by \citet{andrieu2008tutorial} and references therein.

Instead, our objective is to bring the concept of free energy biasing
to the realm of SMC. Specifically, and starting from some pre-specified
sequence $\left(\pi_{t}\right)$, we design a class of SMC samplers,
which compute sequentially the free energy $A_{t}$ associated to
each distribution $\pi_{t}$, and track the sequence of biased densities
$\tilde{\pi}_{t}(\theta)=\pi_{t}(\theta)\exp\left\{ A_{t}\circ\xi(\theta)\right\} $.
In this way, particles may move freely between the modal regions not
only at the early iterations, where $\pi_{t}$ remains close to $\pi_{0}$
and therefore is not strongly multimodal, but also at the later stages,
thanks to free energy biasing. 

We apply free energy SMC sampling to the Bayesian analysis of mixture
models. \citet{ChoLelSto} show that free energy biasing methods are
an interesting approach for dealing with the multimodality of mixture
posterior distributions. In particular, they present several efficient
reaction coordinates for univariate Gaussian mixtures, such as the
hyper-parameter that determines the prior expectation of the component
variances. In this paper, we investigate how free energy SMC compares
with this initial approach based on Adaptive MCMC, and to which extent
such ideas may be extended to other mixture models, such as a bivariate
Gaussian mixture model. 

The paper is organised as follows. Section \ref{sec:SMC} describes
the SMC methodology. Section \ref{sec:FreeEnergy} presents the concept
of free energy biased sampling. Section \ref{sec:FESMC} presents
a new class of SMC methods, termed as free energy SMC. Section \ref{sec:Application-to-mixtures}
discusses the application to Bayesian inference for mixtures, and
presents numerical results, for two types of mixtures (univariate
Gaussian, bivariate Gaussian), and two datasets. Section \ref{sec:Conclusion}
concludes.

\section{SMC algorithms}\label{sec:SMC}

\subsection{Basic structure}

In this section, we describe briefly the structure of SMC algorithms.
For the sake of exposition, we consider a sequence of probability densities
$\pi_{t}$, $t=0,\ldots,T$ defined on a common space $\Theta\subset\mathbb{R}^{d}$.
At each iteration $t$, a SMC algorithm produces a weighted sample
$(w_{t,n},\theta_{t,n})$, $n=1,\ldots,N$, which targets $\pi_{t}$
in the following sense:
\[
\frac{\sum_{n=1}^{N}w_{t,n}\varphi(\theta_{t,n})}{\sum_{n=1}^{N}w_{t,n}}
\rightarrow_{N\rightarrow+\infty}\mathbb{E}^{\pi_{t}}\left\{
  \varphi(\theta)\right\} ,
\]
almost surely, for a certain class of test functions $\varphi$. At
iteration $0$, one typically samples $\theta_{0,n}\sim\pi_{0}$,
and set $w_{0,n}=1$. To progress from iteration $t-1$ to iteration
$t$, it is sufficient to perform a basic importance sampling step
from $\pi_{t-1}$ to $\pi_{t}$: \[
\theta_{t,n}=\theta_{t-1,n},\quad w_{t,n}=w_{t-1,n}\times u_{t}(\theta_{t,n})\]
 where $u_t$ denotes the incremental weight function
\[
u_{t}(\theta)=\frac{\pi_{t}(\theta)}{\pi_{t-1}(\theta)}.\] However, if
only importance sampling steps are performed, the algorithm is
equivalent to a single importance sampling step, from $\pi_{0}$ to
$\pi_{T}$. This is likely to be very inefficient. Instead, one should
perform regularly resample-move steps \citep{GilksBerzu}, that, is, a
succession of i) a resampling step, where current points
$\theta_{t,n}$ are resampled according to their weights, so that
points with a small (resp. big) weight are likely to die
(resp. generate many offsprings); and ii) a mutation step, where each
resampled point is {}``mutated'' according to some probability kernel
$K_{t}(\theta,d\hat{\theta})$, typically a MCMC kernel with invariant
distribution $\pi_{t}$. In the more general formalism of
\citet{DelDouJas:SMC}, this is equivalent to performing an importance
sampling step in the space $\Theta\times\Theta$, with forward kernel
$K_t$, associated to some probability density
$K_t(\theta,\hat\theta)$, and backward kernel $L_{t}$ associated to
the probability density
$L_t(\hat{\theta},\theta)=\pi_{t}(\theta)K_t(\theta,\hat{\theta})/\pi_{t}(\hat{\theta})$.

Resample-move steps should be performed whenever the weight degeneracy
is too high. A popular criterion is EF$(t)<\tau $, where $\tau\in(0,1)$,
and EF is the efficency factor, that is the effective sample size of
\cite{KongLiuWong} divided by $N$, 
\beqn
\mathrm{EF}(t)=\frac{\left(\sum_{n=1}^{N}w_{t,n}\right)^{2}}{N\sum_{n=1}^{N}w_{t,n}^{2}}.
\eeqn

We summarise in Algorithm 1  the general structure of SMC algorithms. 
There are several methods for resampling the particles, e.g. 
the multinomial scheme \citep{Gordon}, the residual scheme 
\citep{LiuChen}, the systematic scheme \citep{Whitley,CarClifFearn}. 
We shall use the systematic scheme in our simulations. 

\begin{flushleft}
\begin{algorithm} \label{algo:SMC}

\caption{A generic  SMC algorithm}

0. Sample $\theta_{0,n}\sim\pi_{0}$, set $w_{0,n}=1$, for $n=1,\ldots,N$.
Set $t=1$. 

1. Compute new weights as \[
w_{t,n}=w_{t-1,n}\times u_{t}(\theta_{t-1,n}).\]

2. If EF$(t)<\tau$, then

(a) resample the particles, i.e. construct a sample
$(\hat\theta_{t,n})_{1\leq n \leq N}$ made of $R_{t,n}$ replicates of
particle $\theta_{t,n}$, $1\leq n \leq N$, where $R_{t,n}$ is a
nonnegative integer-valued random variable such that
\beqn
\mathbb{E} \left[ R_{t,n}
\right]
=   \frac{N w_{t,n}}{\sum_{n'=1}^N w_{t,n'}},
\eeqn 
and set $w_{t,n}=1$.

(b) move the particles with respect to Markov kernel $K_{t}$, \[
\theta_{t,n}\sim K_{t}(\hat{\theta}_{t,n},d\theta)\]

otherwise \[
\theta_{t,n}=\theta_{t-1,n}.\]

3. $t\leftarrow t+1$, if $t<T$ go to Step 1. 
\end{algorithm}

\par\end{flushleft}

\subsection{Adaptiveness of SMC}
\label{sec:adaptivess-smc}

In contrast to MCMC, where designing adaptive algorithms require a
careful convergence study, designing adaptive SMC samplers is
straightforward.  We consider first the design of the MCMC kernels
$K_{t}$. For instance, \citet{Chopin:IBIS} uses independent
Hastings-Metropolis kernels, with a Gaussian proposal fitted to the
current particle sample. This is a reasonable strategy if $\pi_{t}$ is
close to Gaussianity. In this paper, we consider instead the following
strategy, which seems more generally applicable: take $K_{t}$ as a
succession of $k$ Gaussian random walk Hastings-Metropolis steps
$K_{t,i}(\theta,d\theta')$, i.e. simulating from
$K_{t,i}(\theta,d\theta')$ consists of proposing a value $\theta'\sim
N_{d}(\theta,\Sigma_{t,i})$, accepting this value with probability
$1\wedge \left\{ \pi(\theta')/\pi(\theta) \right\} $, otherwise keep
the current value $\theta$. Then take $\Sigma_{t,i}=c_{t,i}S_{t}$,
$c_{t,i}>0$, and $S_{t}$ is the empirical covariance matrix of the
resampled particles at iteration $t$ (that is, the particles obtained
immediately before the MCMC step with kernel $K_{t}$ is
performed). The constant $c_{t,i}$ may be tuned automatically as
well. For instance, one may start with $c_{0}=0.3$, then, each time
the acceptance rate of the MCMC step is below (resp. above) a given
threshold, the constant $c_{t}$ is divided (resp. multiplied) by two.

As in MCMC, it is common to focus on the adaptiveness of the transition
kernels, but one may use the particle sample (or the history of the
process in the MCMC context) to adapt the target distributions as
well. This is precisely what we do in this paper, since the target
at time $t$ on our free energy SMC sampler shall depend on a bias function
which is estimated from the current particle sample, see Section~\ref{sec:FESMC}.

\subsection{IBIS versus annealing, choice of $\pi_{0}$}

When the distribution of interest $\pi$ is some Bayesian posterior
density\[
\pi(\theta)=p(\theta|y_{1:D})=\frac{1}{Z}p(\theta)p(y_{1:D}|\theta),\]
where $y_{1:D}$ is a vector of $D$ observations, $p(\theta)$
is the prior density, and $p(y_{1:D}|\theta)$ is the likelihood,
it is of interest to compare the two aforementioned SMC strategy,
namely, 
\begin{enumerate}
\item IBIS, where $T=D$, and $\pi_{t}(\theta)=p(\theta|y_{1:t})$, in
particular, $\pi_{0}(\theta)=p(\theta)$ is the prior; and 
\item Annealing, where $\pi_{t}(\theta)=\pi_{0}(\theta)^{1-\gamma_{t}}\pi(\theta)^{\gamma_{t}}$,
$\gamma_{t}$ is an increasing sequence such that $\gamma_{0}=0$,
and $\gamma_{T}=1$, $\pi_{0}$ is typically the prior density, but
could be something else, and $T$ and $D$ do not need to be related.
\end{enumerate}
Clearly, for the same number of particles, and assuming that the same
number of resample-move steps is performed, IBIS is less time-consuming,
because calculations at iteration $t$ involve only the $t$ first
observations. On the other hand, annealing may produce a smoother
sequence of distributions, so it may require less resample-move steps.
\citet{jasra2007population} provide numerical examples where the
IBIS strategy leads to unstable estimates. In the context discussed
in the paper, see Section \ref{sec:Application-to-mixtures}, and
elsewhere, we did not run into cases where IBIS is particularly unstable.
Perhaps it is fair to say that a general comparison is not meaningful,
as the performance of both strategies seems quite dependent on the
applications, and also various tuning parameters such as the sequence
$\gamma_{t}$ for instance. 

We take this opportunity however to propose a simple method to improve
the regularity of the IBIS sequence, in the specific case where the
observations are exchangeable and real-valued.  We remark first that
this regularity depends strongly on the order of incorporation of the
$y_t$'s. For instance, sorting the observations in ascending order
would certainly lead to very poor performance. On the other hand, a
random order would be more suitable, and was recommended by
\cite{Chopin:IBIS}.  Pushing this idea further, we propose the
following strategy: First, we re-define the median of a sample as
either the usual median, when $D$ is an odd number, or the smallest of
the two middle values in the ordered sample, when $D$ is an even
number. Then, we take $y_{1}$ as the median observation, $y_{2}$
(resp. $y_{3}$) to be the median of the observations that are smaller
(resp. larger) than $y_{1}$, then we split again the four
corresponding sub-samples by selecting some values $y_{4}$ to $y_{7}$,
and so on, until all values are selected. We term this strategy as
``Van der Corput ordering'', as a Van der Corput binary sequence is
precisely defined as $1/2,$ $1/4,$ $3/4$, $1/8,\ldots$

A point which is often overlooked in the literature, and which affects
both strategies, is the choice of $\pi_{0}$. Clearly, if
$\pi_{0}(\theta)=p(\theta)$, one may take the prior so uninformative
that the algorithm degenerates in one step. Fortunately, in the
application we discuss in this paper, namely Bayesian analysis of
mixture models, priors are often informative; see Section
\ref{sec:Application-to-mixtures} for a discussion of this point. In
other contexts, it may be helpful to perform a preliminary exploration
of $\pi$ in order design some $\pi_{0}$, quite possibly different from
the prior, so that (i) for the annealing strategy, moving from
$\pi_{0}$ to $\pi_{T}=\pi$ does not take too much time; and (ii) for
the IBIS strategy, one can use $\pi_{0}$ as an artificial prior, and
recover the prior of interest at the final stage of the algorithm, by
multiplying all the particle weights by $p(\theta)/\pi_{0}(\theta)$.

\section{Free energy-biased sampling}\label{sec:FreeEnergy}

\subsection{Definition of free energy, and free-energy biased densities}

In this section we explain in more detail the concept of free energy
biasing. We consider a single distribution of interest, defined by
a probability density $\pi$ with respect to the Lebesgue measure
associated to $\Theta\subset\mathbb{R}^{d}$. As explained in the
introduction, the first step in implementing a free energy biasing
method is to choose a reaction coordinate, that is, some measurable
function $\xi:\theta\rightarrow\mathbb{R}^{d'}$, where $d'$ is small.
In this paper, we take $d'=1.$ One assumes that the multimodality
of $\pi$ is strongly determined, in some sense, by the direction
$\xi(\theta)$. For instance, the distribution of $\theta$, conditional
on $\xi(\theta)=x$, may be much less multimodal than the complete
distribution $\pi$, for either all or certain values of $x$. 

In words, the free energy is, up to an arbitrary constant, minus the
logarithm of the marginal density of $\xi(\theta)$. The free energy
may be written down informally as
\beqn
\exp\left\{ -A(x)\right\}
\propto\int_{\Theta}\pi(\theta)\I_{[x,x+dx]}\left\{
  \xi(\theta)\right\} \, d\theta
\eeqn
and more rigorously, as 
\[
\exp\left\{ -A(x)\right\} \propto\int_{\Omega_{x}}\pi(\theta)\, d\left\{ \theta|\xi(\theta)=x\right\} ,\]
where $\Omega_{x}=\left\{ \theta\in\Theta:\,\xi(\theta)=x\right\} $,
and $d\left\{ \theta|\xi(\theta)=x\right\} $ denotes the conditional
measure on the set $\Omega_{x}$ which is ``compatible'' with Lebesgue
measure on the embedding space  $\Theta$, i.e. volumes are preserved
and so on. In both formulations, the proportionality
relation indicates that the density $\pi$ may be known only up to
a multiplicative constant, and therefore that the free energy is defined
only up to an arbitrary additive constant. 

The free energy biased density $\tilde{\pi}$ is usually  defined as 
\beqn \label{eq:febpi}
\tilde{\pi}(\theta)\propto\pi(\theta)\exp\left\{
  A\circ\xi(\theta)\right\} 
\I_{[x_{\min},x_{\max}]}\left\{ \xi(\theta)\right\} 
\eeqn
where $[x_{\min},x_{\max}]$ is some pre-defined range. It is clear
that, with respect to $\tilde{\pi}$, the marginal distribution of
the random variable $\xi(\theta)$ is uniform over $[x_{\min},x_{\max}]$,
and the conditional distributions of $\theta$, given $\xi(\theta)=x$
matches the same conditional distribution corresponding to $\tilde{\pi}$.
The objective is to sample from $\tilde{\pi}$, which requires to
estimate the free energy $A$. 

To avoid the truncation incurred by the restriction to interval
$[x_{\min},x_{\max}]$, we shall consider instead the
following version of the free-energy biased density $\tilde{\pi}_t$: 
\beqn \label{eq:febpi2}
\tilde{\pi}(\theta)\propto\pi(\theta)\exp\left\{
  A\circ\xi(\theta)\right\}
\eeqn
where the definition of $A$ is extended as follows: $A(x)=A(x_{\min})$
for $x\leq x_{\min}$, $A(x)=A(x_{\max})$ for $x\geq x_{\max}$. 

\subsection{Estimation of the free energy}\label{sub:EstimFE}

As explained in the introduction, one usually resorts to some form
of Adaptive MCMC to estimate the free energy $A$. Specifically, one
performs successive MCMC steps (typically Hastings-Metropolis), such
that the Markov kernel $K_{(t)}$ used at iteration $t$ depends on
the trajectory of the simulated process up to time $t-1$. (The simulated
process is therefore non-Markovian.) The invariant distribution of kernel $K_{(t)}$ is
 $\pi_{(t)}(\theta)\propto\pi(\theta)\exp\left\{ A_{(t)}\circ\xi(\theta)\right\} $,
where $A_{(t)}$ is an estimate of the free energy $A$ that has been
computed at iteration $t$, from the simulated trajectory up to time
$t-1$. Note that the brackets in the notations $K_{(t)}$, $\pi_{(t)}$,
$A_{(t)}$ indicate that all these quantities are specific to this
section and to the Adaptive MCMC context, and must not mistaken for
the similar quantities found elsewhere in the paper, such as, e.g.
the density $\pi_{t}$ targeted at iteration $t$ by a SMC sampler.
The difficulty is then to come up with an efficient estimator (or
rather a sequence of estimators, $A_{(t)}$), of the free energy. 

Since this paper is not concerned with adaptive MCMC, we consider
instead the much simpler problem of estimating the free energy $A$
from a weighted sample $(\theta_{n},w_{n})_{n=1,\ldots,N}$ targeting
$\pi$; for instance, the $\theta_{n}$'s could be i.i.d. with probability
density $g$, and $w_{n}=\pi(\theta)/g(\theta)$. Of course, this
discussion is simplistic from an Adaptive MCMC perspective, but it will
be sufficient in our SMC context. We refer the reader to e.g.
\cite{ChoLelSto} for the missing details. 

First, it is necessary to discretise the problem, and consider some
partition:
\begin{equation}
[x_{\min},x_{\max}]=\cup_{i=0}^{n_{x}}[x_{i},x_{i+1}],\quad
x_{i}=x_{\min}+(x_{\max}-x_{\min})\frac{i}{n_{x}}.
\label{eq:grid}
\end{equation}

Then, they are basically two ways to estimate $A$. The first method
is to estimate directly a discretised version of $A$, by simply computing
an estimate the proportion of points that fall in each bin: 
\[
\exp \left\{ -\hat{A}_{1}(x)  \right\}=\frac{\sum_{n=1}^{N}w_{n}\I\left\{ \xi(\theta_{n})
\in[x_{i},x_{i+1}]\right\} }{\sum_{n=1}^{N}w_{n}},\quad\mbox{for
}x\in[x_{i},x_{i+1}].
\]
The second method is indirect, and based on the following property:
the derivative of the free energy is such that
 \[
A'(x)=\mathbb{E}^{\pi}\left[f(\theta)|\xi(\theta)=x\right]\]
where the force $f$ is defined as:

\[
f=-\frac{\left(\nabla\log\pi\right)\cdot\left(\nabla\xi\right)}{|\nabla\xi|^{2}}-\mathrm{div}\left(\frac{\nabla\xi}{|\nabla\xi|^{2}}\right),\]
and $\nabla$ (resp. $\mathrm{div}$) is the gradient (resp. divergence)
operator. Often, $\xi(\theta)$ is simply a coordinate of the vector
$\theta$, $\theta=(\xi,...)$, in which case the expression above
simplifies to $f=-\partial\log\pi/\partial\xi$. This leads to the
following estimator of the derivative of $A$:

\[
\hat{A}'_{2}(x)=\frac{\sum_{n=1}^{N}w_{n}\I\left\{ \xi(\theta_{n})\in[x_{i},x_{i+1}]\right\} f(\theta_{n})}{\sum_{n=1}^{N}w_{n}\I\left\{ \xi(\theta_{n})\in[x_{i},x_{i+1}]\right\} },\quad\mbox{for }x\in[x_{i},x_{i+1}].\]
Then an estimate of $A$ may be deduced by simply computing cumulative
sums for instance:\[
\hat{A}_{2}(x)=\sum_{j:x_{j}\leq x}\hat{A'}_{2}(x_{j})(x_{j+1}-x_{j}),\quad\mbox{for }x\in[x_{i},x_{i+1}].\]

Methods based on the first type of estimates are usually called ABP
(Adaptive Biasing Potential) methods, while methods of the second
type are called ABF (Adaptive Biasing Force). Empirical evidence suggests
that ABF leads to slightly smoother estimates, presumably because
it is based on a derivative.

\section{Free energy SMC}\label{sec:FESMC}

We now return to the SMC context, and consider a pre-specified sequence
$(\pi_{t})$. Our objective is to derive a SMC algorithm which sequentially
compute the free energy $A_{t}$ associated to each density $\pi_{t}$,
\[
\exp\left\{ - A_t(x)\right\} \propto\int\pi_{t}(\theta)d\left\{
  \theta|\xi(\theta)=x\right\} \] and sample $\tilde{\pi}_{t}$, the
free energy biased version of $\pi_{t}$,
 \beqn \label{eq:febpit}
\tilde{\pi}_t(\theta)\propto\pi_t(\theta) \exp\left\{
  A_t\circ\xi(\theta)\right\}.  \eeqn 
Again, to avoid truncating to
interval $[x_{\min},x_{\max}]$, one extends the definition of $A_t$
outside $[x_{\min},x_{\max}]$ by taking $A_{t}(x)=A_t(x_{\min})$ for
$x<x_{\min}$, $A_{t}(x)=A_t(x_{\max})$ for $x>x_{\max}$.

As explained in Section \ref{sub:EstimFE}, one actually estimates
a discretised version of the free energy, i.e., the algorithm shall
provide estimates $\hat{A}_{t}(x_{i})$, $i=0,\ldots,n_{x}$ of the
free energy evaluated at grid points over an interval $[x_{\min},x_{\text{\ensuremath{\max}}}]$,
as defined in \eqref{eq:grid}. Note that this grid is the same for all
iterations $t$. 

Assume that we are at the end of iteration $t-1$, that estimates
$\hat{A}_{t-1}(x_{i})$ of $A_{t-1}$ have been obtained, and that
the particle system $(\theta_{t-1,n},w_{t-1,n})_{n=1,\ldots,N}$ targets
$\tilde{\pi}_{t-1}$. If the particles are re-weighted according to the incremental weight
function $u_{t}(\theta)=\pi_{t}(\theta)/\pi_{t-1}(\theta)$, i.e.
\[
\bar{w}_{t,n}=w_{t-1,n}\times u_{t}(\theta_{t-1,n})\]
then the new target distribution of the particle system $(\theta_{t-1,n},\bar{w}_{t,n})_{n=1,\ldots,N}$
is \[
\bar{\pi}_{t}(\theta)\propto\tilde{\pi}_{t-1}(\theta)u_{t}(\theta).\]
The objective is then to recover $\tilde{\pi}_t$, which depends
on the currently unknown free energy $A_t$. To that effect, we
first state the following result. 
\bteo
The free energy $D_t$ associated to $\bar{\pi}_{t}$ is 
\[
D_{t}=A_{t}-A_{t-1}\]
that is, the difference between the free energies of $\pi_{t}$
and $\pi_{t-1}$. 
\eteo
\bpro 
One has, for $\theta\in\Theta$,
 \[
\bar{\pi}_{t}(\theta)\propto\pi_t(\theta)\exp\left\{ A_{t-1}\circ\xi(\theta)\right\} \]
hence, for $x\in\xi(\Theta)$,
 \[
\int_{\Omega_{x}}\bar{\pi}_t(\theta)d\left\{ \theta|\xi(\theta)=x\right\} =\exp\left\{ \left(A_{t-1}-A_t\right)(x)\right\} .\]
and one concludes. \epro

This result provides the justification for the following strategy.
First, particles are reweighted from $\pi_{t-1}$ to $\bar{\pi}_{t}$,
as explained above. Second, the free energy $D_{t}$ of $\bar{\pi}_{t}$
is estimated, using either the ABP or the ABF strategy, see Section
\ref{sub:EstimFE}; this leads to some estimate $\hat{D}_{t}$ of
$D_{t}$, ot more precisely estimates $\hat{D}_{t}(x_{i}$) over the
grid $x_{0},\ldots,x_{n_{x}}$. From this, one readily obtains estimates
of the current free energy, using the proposition above:
 \beqnn \label{eq:updateAt}
\hat{A}_{t}(x_{i})=\hat{A}_{t-1}(x_{i})+\hat{D}_{t}(x_{i}),\quad
i=0,\ldots,n_{x}.
\eeqnn
Third, one recovers $\tilde{\pi}_{t}$ by performing an importance
sampling step from $\bar{\pi}_{t}$ to $\tilde{\pi}_{t}$; this is
equivalent to updating the weights as follows:
\[
w_{t,n}=\bar{w}_{t,n}\exp\left\{ \hat{D_{t}}\circ\xi(\theta_{t,n})\right\} .\]
An outline of this free energy SMC algorithm is given in Algorithm 2.

\begin{flushleft}
\begin{algorithm} \label{algo:feSMC}

\caption{Free energy  SMC}

0. Sample $\theta_{0,n}\sim\pi_{0}$, set $w_{0,n}=1$, for
$n=1,\ldots,N$. Compute $A_0$ and set $t=1$. 

1. Compute new weights as \[
\bar{w}_{t,n}=w_{t-1,n}\times u_{t}(\theta_{t-1,n}).\]

2. Compute an estimator $\hat D_t$ of free energy $D_t$,  compute weights
\beqn
w_{t,n} = \bar{w}_{t,n} \exp \left[ \hat {D}_t\circ\xi
  (\theta_{t-1,n}) \right]
\eeqn
and update the estimate $\hat A_t$ of the free energy $A_t$, using~\eqref{eq:updateAt}.

3. If EF$(t)<\tau$, then

(a) resample the particles, i.e. draw randomly $\hat{\theta}_{t,n}$
in such a way that 
\beqn
\mathbb{E} \left[ \sum_{n'=1}^N \I(\hat\theta_{t,n'}=\theta_{t-1,n})
  \bg (\theta_{t-1,n},w_{t,n})
\right]
=   \frac{N w_{t,n}}{\sum_{n'=1}^N w_{t,n'}}
\eeqn 
and set $w_{t,n}=1$.

(b) move the particles with respect to Markov kernel $K_{t}$, \[
\theta_{t,n}\sim K_{t}(\hat{\theta}_{t,n},d\theta)\]

otherwise \[
\theta_{t,n}=\theta_{t-1,n}.\]

4. $t\leftarrow t+1$, if $t<T$ go to Step 1. 
\end{algorithm}

\par\end{flushleft}

At the final stage of the algorithm (iteration $T$), one recovers
the unbiased target $\pi_{T}=\pi$ by a direct importance sampling
step, from $\tilde{\pi}_{T}$ to $\pi_{T}$: \[
\frac{\pi_{T}(\theta)}{\tilde{\pi}_{T}(\theta)}\propto\exp\left\{ \hat{A}_{T}\circ\xi(\theta)\right\} .\]
This is because of this ultimate debiasing step, which relies on $\hat{A}_{T}$,
that one must store in memory and compute iteratively the {}``complete''
free energy $A_{T}$ (as opposed to the successive $D_{t}$, which
may be termed as {}``incremental'' free energies). If this unbiasing
step is too {}``brutal'', meaning that too many particles get a
low weight in the final sample, then one may apply instead a progressive
unbiasing strategy, by extending the sequence of distributions $\tilde{\pi}_{T}$
as follows:
 \[
\tilde{\pi}_{T+l}(\theta)\propto\tilde{\pi}_{T}(\theta)\exp
\left\{ \left(\frac{l}{L}\right)\hat{A}_{T}\circ\xi(\theta)\right\} ,\quad l=0,\ldots,L\]
and performing additional SMC steps, that is, successive importance
sampling steps from $\tilde{\pi}_{T+l}$ to $\tilde{\pi}_{T+l+1}$,
and, when necessary, resample-move steps in order to avoid degeneracy.
In our simulations, we found that progressive unbiasing did lead to
some improvement, but that often direct unbiasing was
sufficient. Hence, we report only results from direct unbiasing in
the next Section.

\section{Application to mixtures}\label{sec:Application-to-mixtures}

\subsection{General formulation, multimodality}

A $K-$component Bayesian mixture model consists of $D$ independent
and identically distributed observations $y_{i}$, with parametric
density \[
p(y_{i}|\theta)=\frac{1}{\sum_{k=1}^{K}\omega_{k}}\sum_{k=1}^{K}\omega_{k}\psi(y_{i};\xi_{k}),\quad\omega_{k}\geq0,\]
 where $\left\{ \psi(\cdot;\xi),\,\xi\in\Xi\right\} $ is some parametric
family, e.g. $\psi(y,\xi)=N(y;\mu,1/\lambda)$, $\xi=(\mu,\lambda^{-1})$.
The parameter vector contains
 \[
\theta=(\omega_{1},\ldots,\omega_{k},\xi_{1},\ldots,\xi_{k},\eta),\]
where $\eta$ is the set of hyper-parameters that are shared by the
$K$ components. The prior distribution $p(\theta)$ is typically
 symmetric with respect to component permutation. In particular,
one may assume that, a priori and independently $\omega_{k}\sim\mathrm{Gamma}(\delta,1)$.
This leads to a Dirichlet$_{K}(\delta,\ldots,\delta)$ prior for the
component probabilities
\[
q_{k}= \frac{\omega_{k}}{\sum_{l=1}^{K}\omega_{l}},\quad k=1,\ldots K.\]
We note in passing that, while the formulation of a mixture model
in terms of the $q_{k}'s$ is more common, we find that the formulation
in terms of the unnormalised weights $\omega_{k}$ is both more tractable
(because it imposes symmetry in the notations) and more convenient
in terms of implementation (e.g. designing Hastings-Metropolis steps). 

An important feature of the corresponding posterior density
 \[
\pi(\theta)=p(\theta|y_{1:D})\propto p(\theta)\prod_{i=1}^{D}p(y_{i}|\theta),\]
assuming $D$ observations are available, is its invariance with
respect to {}``label permutation''. This feature and its bearings
to Monte Carlo inference have received a lot of attention, see e.g.
\citet{CelHurRob}, \citet{jasra2005markov}, \citet{ChoLelSto} among
others. In short, a standard MCMC sampler, such as the Gibbs sampler
of \cite{DiebRob94}, see also the book of \citet{Fru:book}, typically visits a
single modal region. But, since the posterior is symmetric, any mode
admits $K!-1$ replicates in $\Theta$. Therefore, one can assert
that the sampler has not converged. \citet{Fru:mixtures} proposes
to permute randomly the components at each iteration. However, \citet{jasra2005markov}
mentions the risk of ``genuine multimodality'', that is, the $K!$
symetric modal regions visited by the permutation sampler may still
represent a small part of the posterior mass, because other sets of
equivalent modes have not been visited.  \cite[][Chap. 6]{BayesianCore} and \citet{ChoLelSto}
provide practical examples of this phenomenon. 

One could say that random permutations merely ``cure the most obvious
symptom'' of failed convergence. We follow \citet{CelHurRob}, \citet{jasra2005markov}
and \citet{ChoLelSto}, and take the opposite perspective that one
should aim at designing samplers that produce a nearly symmetric output
(with respect to label switching), \emph{without resorting to random
permutations.}

\subsection{Univariate Gaussian mixtures}

\subsubsection{Prior, reaction coordinates }

We first consider a univariate Gaussian mixture model, i.e. $\psi(y,\xi)=N(y;\mu,\lambda^{-1})$,
$\xi=(\mu,\lambda^{-1})$, and we use the same prior as in \citet{RichGreen},
that is, for $k=1,\ldots,K$, independently, 
\[
\mu_{k}\sim N(M,\kappa^{-1}),\quad\lambda_{k}\sim\mathrm{Gamma}(\alpha,\beta),\]
where $\alpha$, $M$ and $\kappa$ are fixed, and $\beta$ is a
hyper-parameter:\[
\beta\sim\mathrm{Gamma}(g,h).\]
Specifically, we take $\delta=1$, $\alpha=2$ (see Chap. 6 of \citealp{Fru:book} for a justification),
$g=0.2$, $h=100g/\alpha R^{2}$, $M=\bar{y}$, and $\kappa=4/R^{2}$,
where $\bar{y}$ and $R$ are, respectively, the empirical mean and
the range of the observed sample. 

Regarding the application of free energy methods to univariate
Gaussian mixture posterior distributions, 
\citet{ChoLelSto} find that the two following functions of $\theta$
are efficient reaction coordinates: $\xi(\theta)=\beta$, and the
potential function $V(\theta)=-\log\left\{ p(\theta)p(y_{1:D}|\theta)\right\} $,
that is, up to a constant, minus log the posterior density. However,
the latter reaction coordinate is less convenient, because it is difficult
to determine in advance the range $[x_{\min},x_{\max}]$ of exploration.
This is even more problematic in our sequential context. Using the
IBIS strategy for instance, one would define $V_{t}(\theta)=-\log\left\{ p(\theta)p(y_{1:t}|\theta)\right\} $,
but the range of likely values for $V_{t}$ would typically be very
different between small and large values of $t$. Thus we discard
this reaction coordinate. 

In constrast, as discussed already in \citet{ChoLelSto}, it is reasonably
easy to determine a range of likely values for the reaction coordinate
$\xi(\theta)=\beta$. In our simulations, we take
$[x_{\min},x_{\max}]=[R^{2}/2000,R^{2}/20]$, where, again,  $R$ is the range of
the data.  \citet{ChoLelSto} explains the good performance of this
particular reaction coordinate as follows. Large values of $\beta$ 
penalise small component variances,  thus forcing $\beta$ to large
values leads to a conditional posterior distribution which favours
overlapping components, which may switch more easily.

\subsubsection{Numerical example}

We consider the most challenging example discussed in \citet{ChoLelSto},
namely the Hidalgo stamps dataset, see e.g. \cite{Izenman}  for details, and
$K=3$. In particular, \citet{ChoLelSto} needed about $10^9$ iterations
of an Adaptive MCMC sampler (namely, an ABF sampler) to obtain a
stable estimate of the free energy. 

We run SMC samplers with the following settings:
the number of particles is $N=2\times 10^4$, the criterion for
triggering resample-move steps is ESS$<0.8 N$, and a move step
consists of  10 successive Gaussian random walk steps, using the
automatic calibration strategy described in Section
\ref{sec:adaptivess-smc}. 

\begin{figure}[H] \label{fig:nobias_trace}
\begin{center}
  \includegraphics[width=0.3\textwidth]{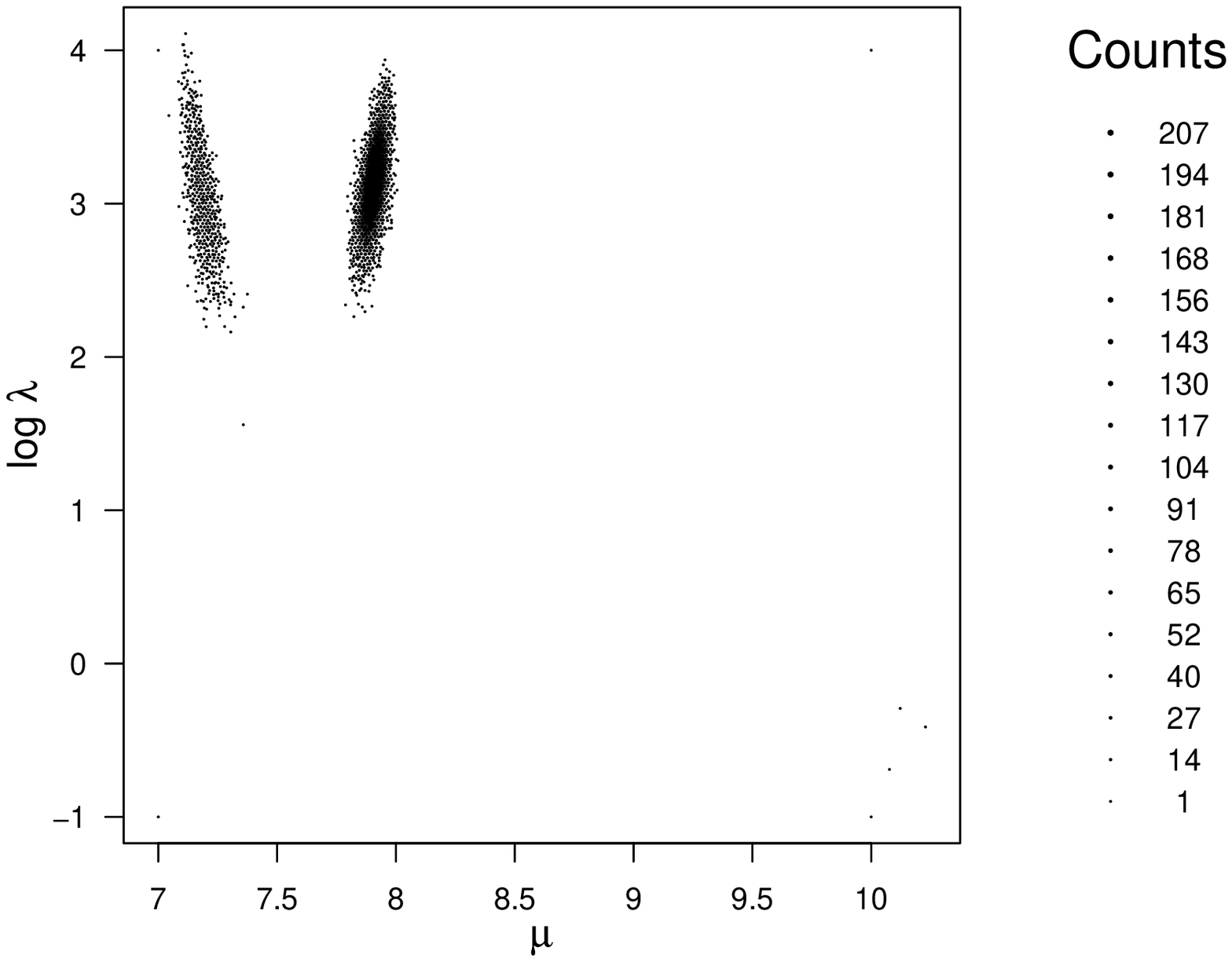}
  \includegraphics[width=0.3\textwidth]{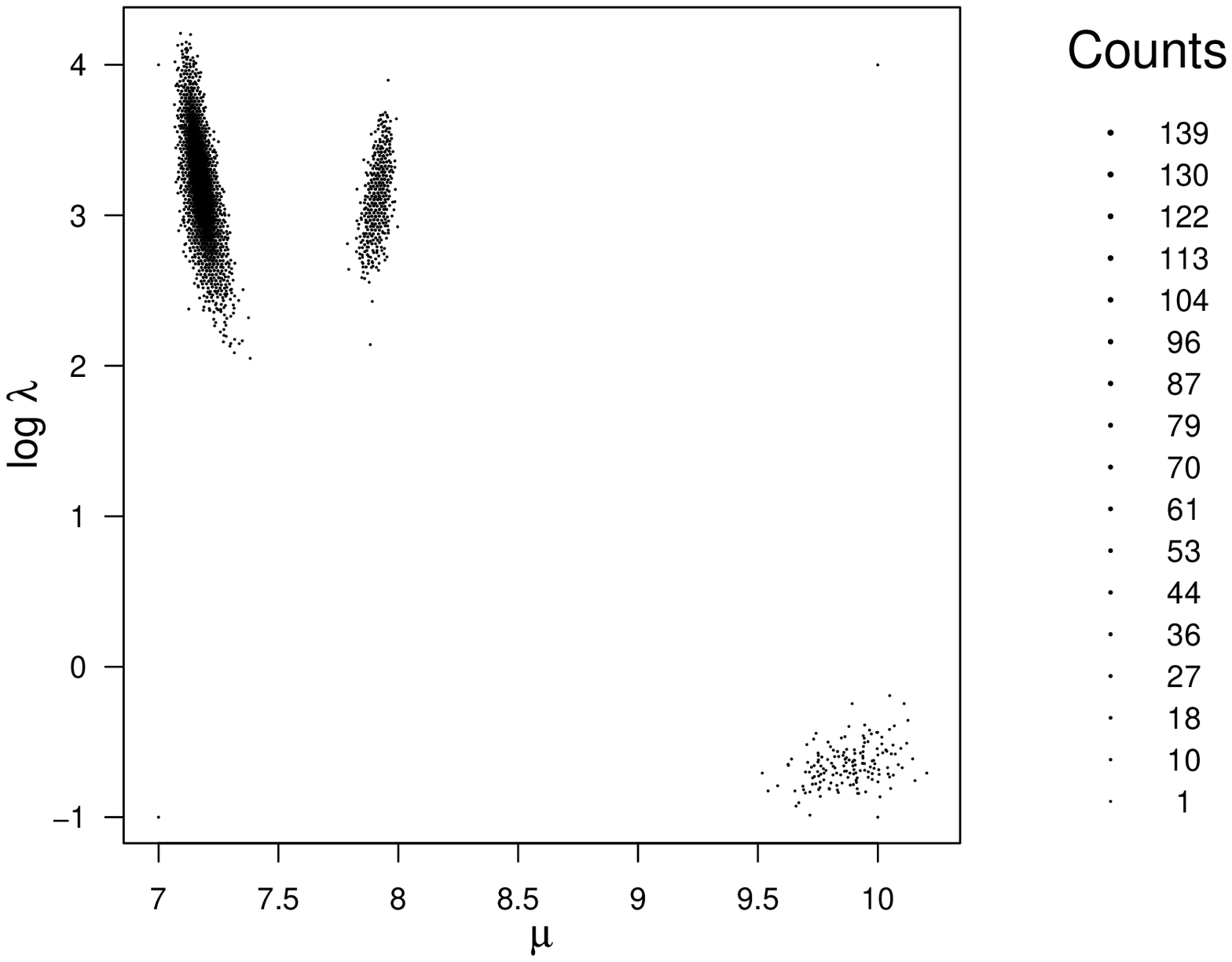}
  \includegraphics[width=0.3\textwidth]{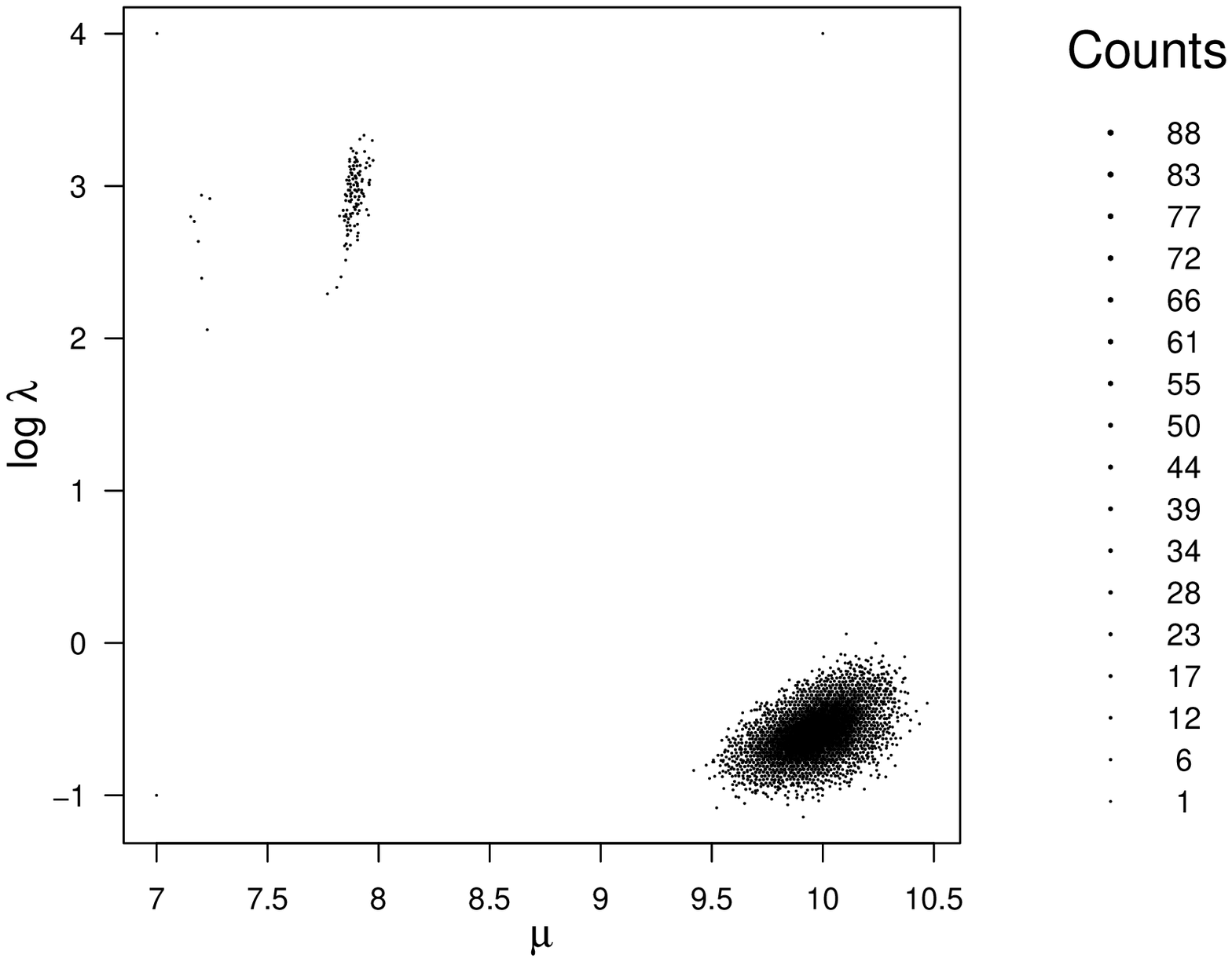}
\end{center}
 \caption{\label{traceplots1} Hexagon binning for
   $(\mu_k,\log\lambda_k)$, $k=1$, 2, 3,  for the standard SMC
   sampler, no free energy biasing, IBIS strategy. }
 \end{figure}

\begin{figure}[H]\label{fig:nobias_hist}
\begin{center}
  \includegraphics[width=0.4\textwidth]{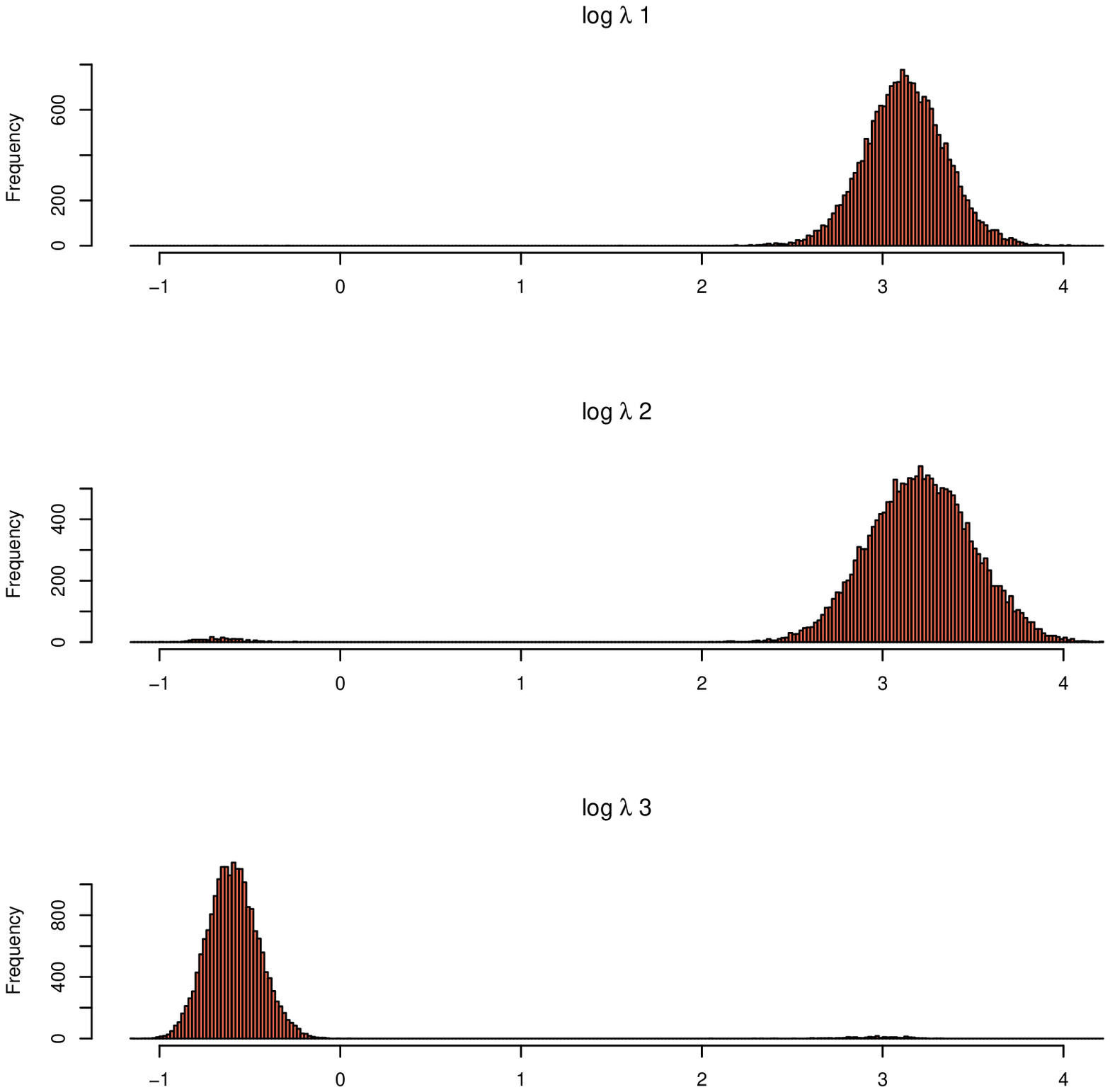}
  \includegraphics[width=0.4\textwidth]{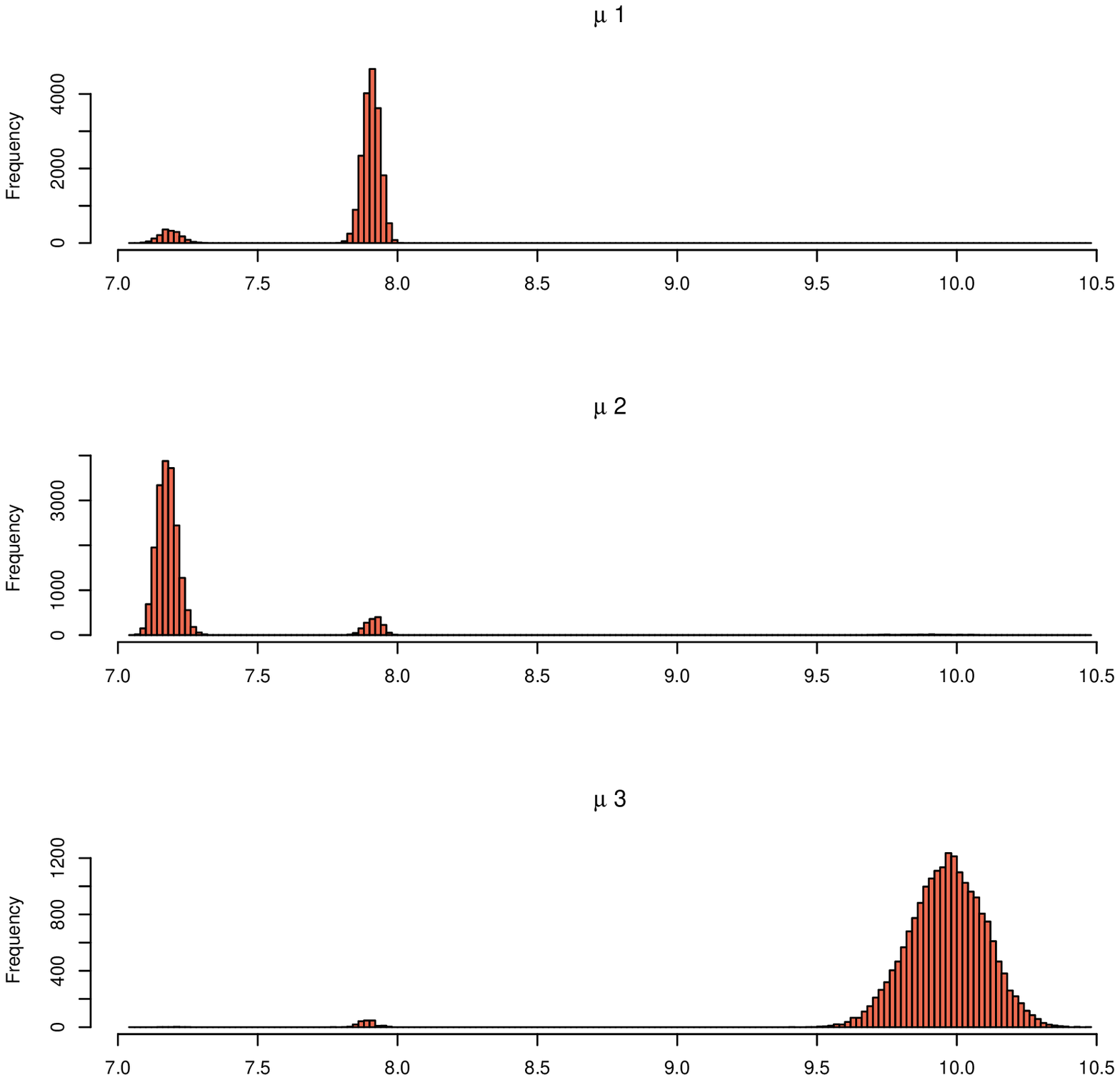}
\end{center}
 \caption{\label{histograms1} Weighted 1D histograms for the standard SMC
   sampler, no free energy biasing, IBIS strategy.}
 \end{figure}

We first run a SMC sampler, without free energy biasing, and using the
IBIS strategy. Results are reported in Figures \ref{traceplots1}
and \ref{histograms1}: the output is not symmetric with
respect to label permutation, and only one modal region  of the posterior 
distribution is visited. 

\begin{figure}[H]
\begin{center}
  \includegraphics[width=0.3\textwidth]{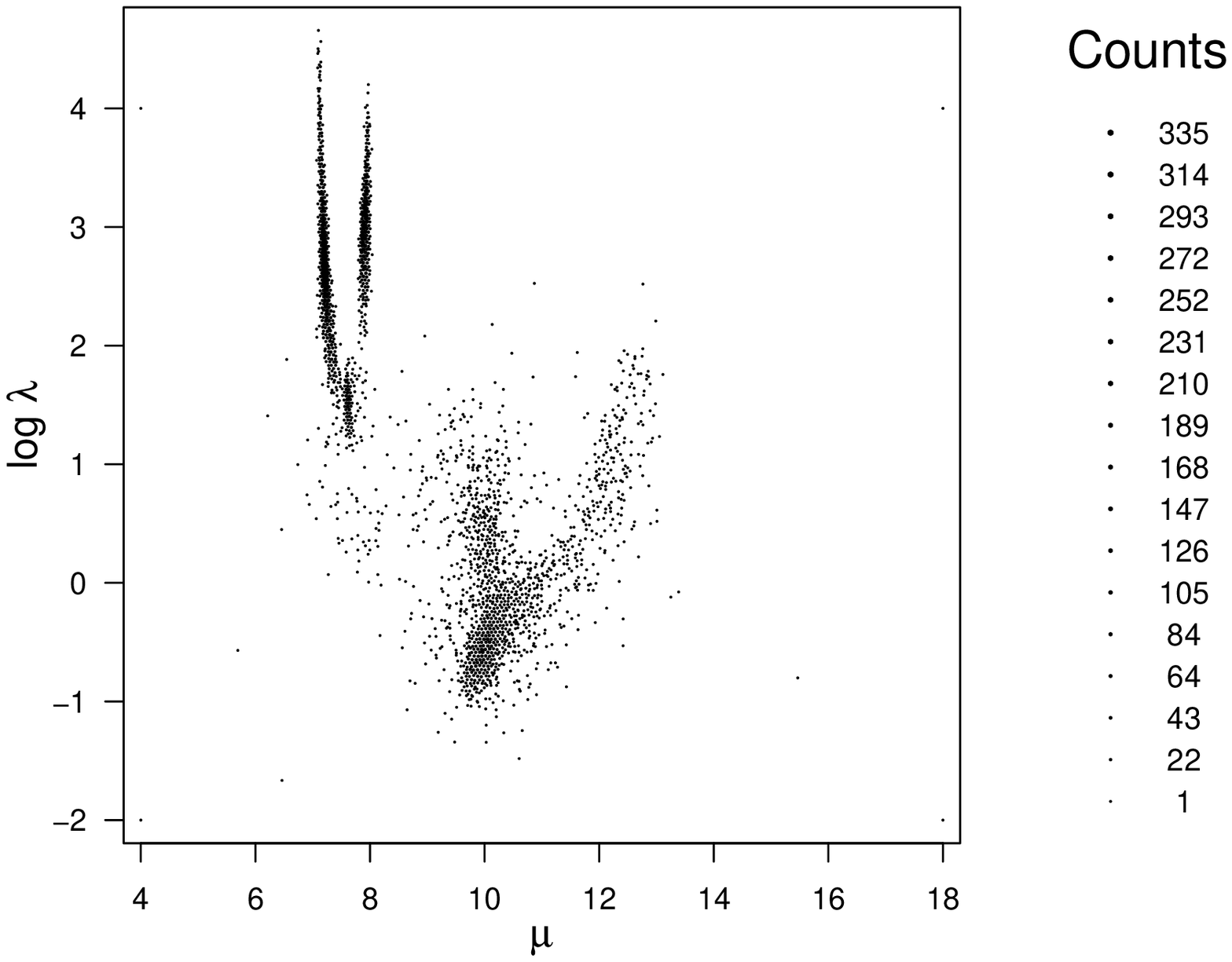}
  \includegraphics[width=0.3\textwidth]{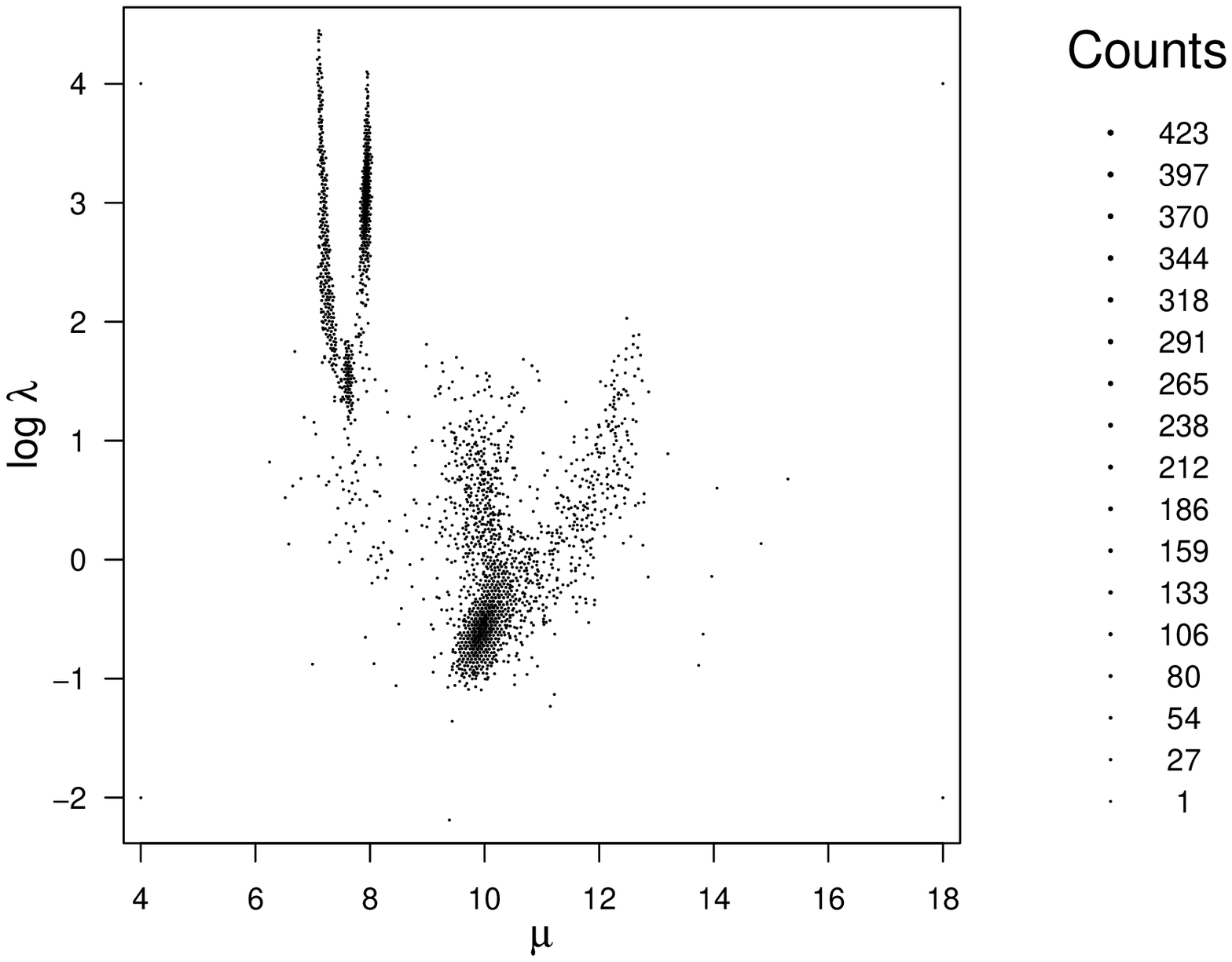}
  \includegraphics[width=0.3\textwidth]{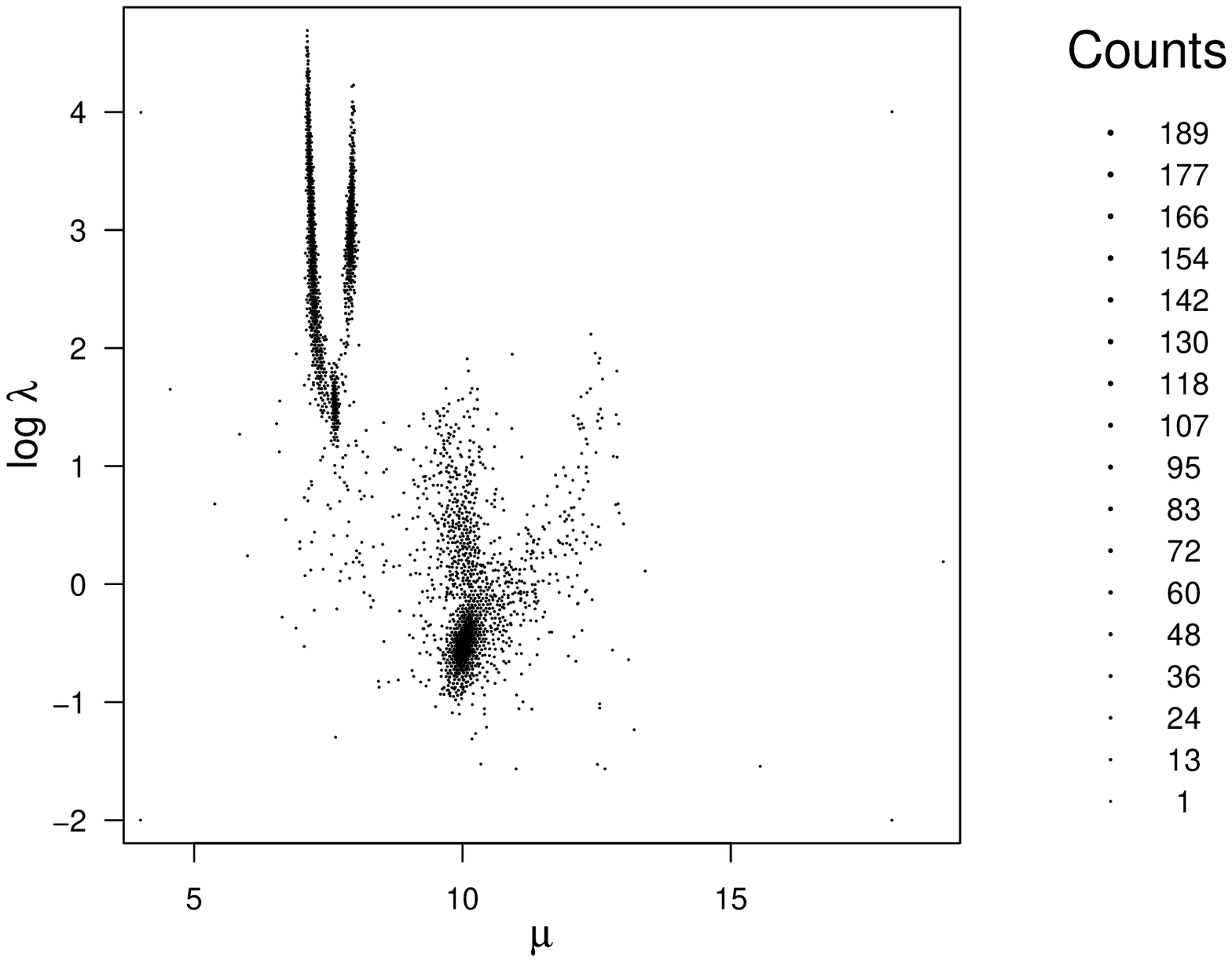}
\end{center}
 \caption{\label{traceplots2}  Hexagon binning for
   $(\mu_k,\log\lambda_k)$, $k=1$, 2, 3,  for the free energy SMC
   sampler,  before the final debiasing step, IBIS strategy.}
 \end{figure}

\begin{figure}[H]
\begin{center}
  \includegraphics[width=0.4\textwidth]{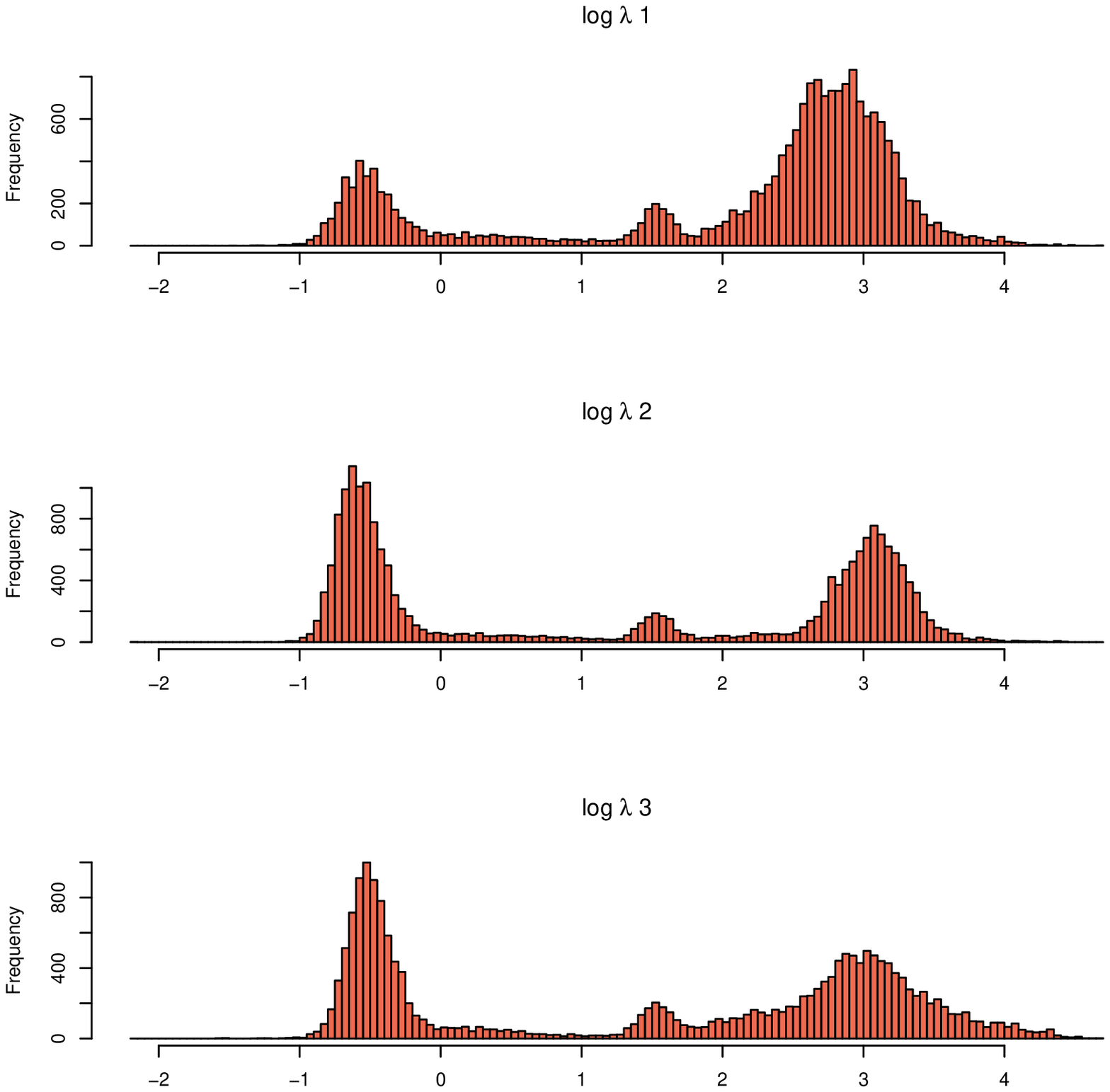}
  \includegraphics[width=0.4\textwidth]{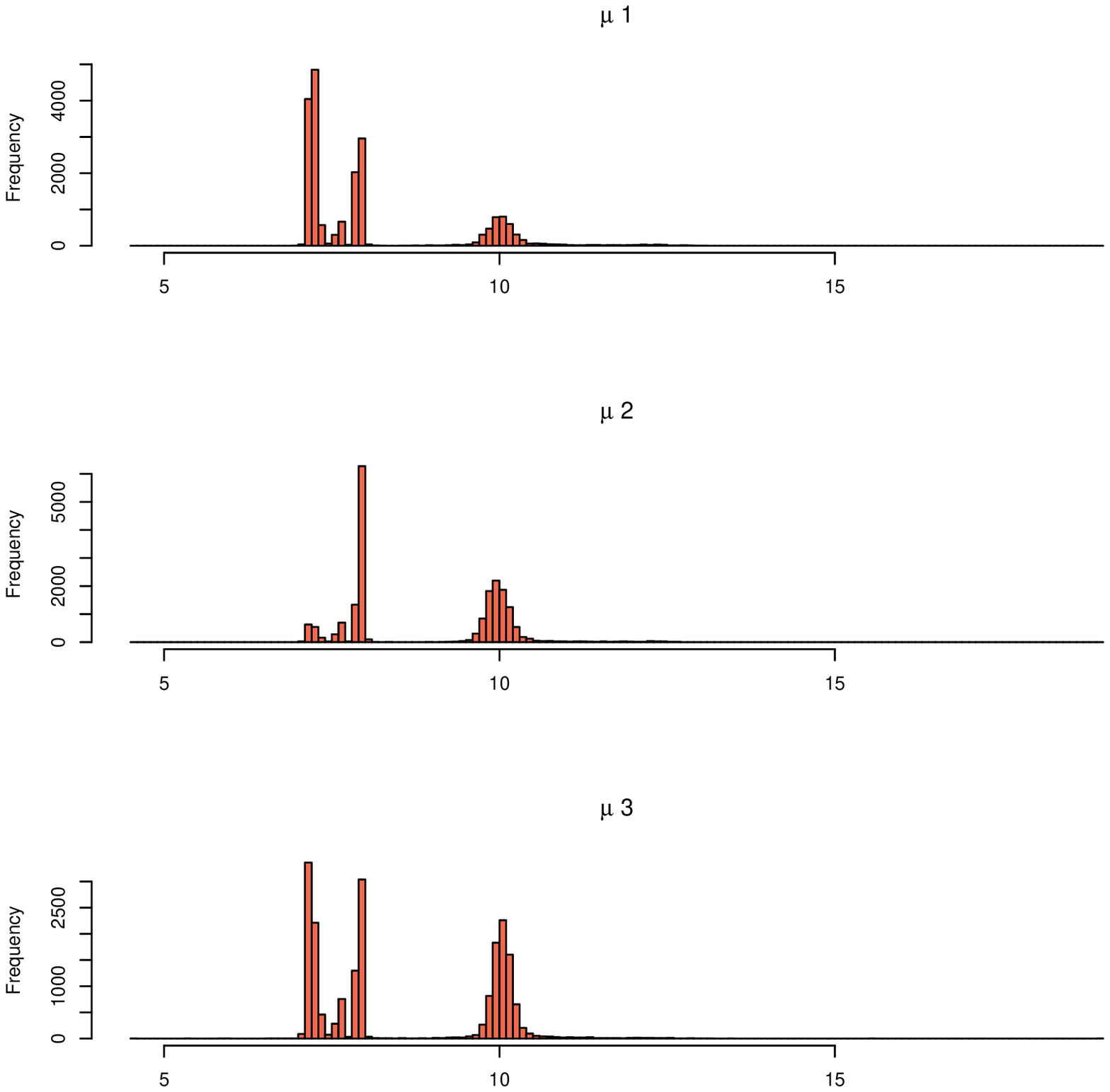}
\end{center}
 \caption{\label{histograms2} Histograms of the components of the
   simulated particles obtained by free energy SMC
   sampler,  before the final debiasing step, IBIS strategy.} 
 \end{figure}

\begin{figure}[H]
\begin{center}
  \includegraphics[width=0.3\textwidth]{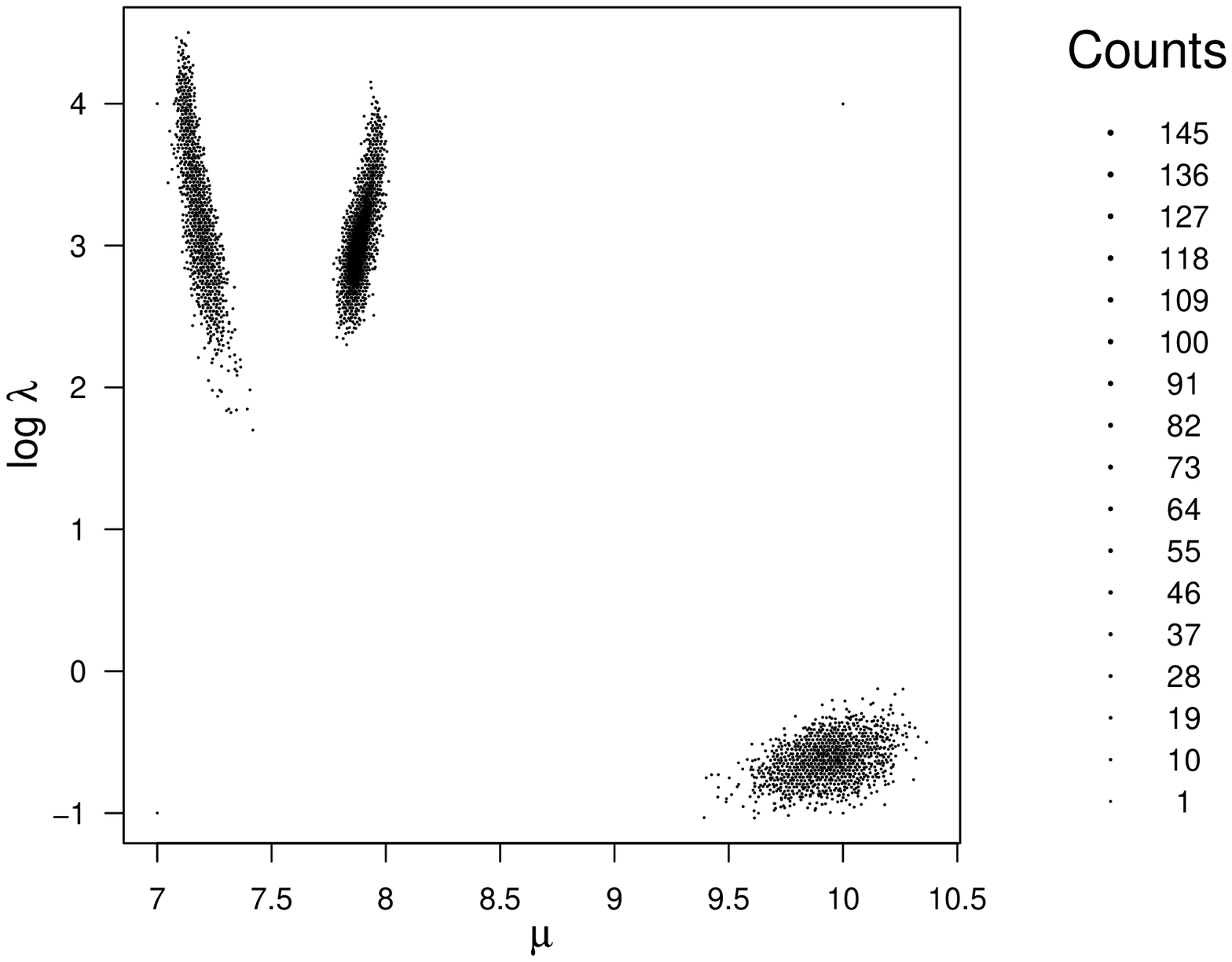}
  \includegraphics[width=0.3\textwidth]{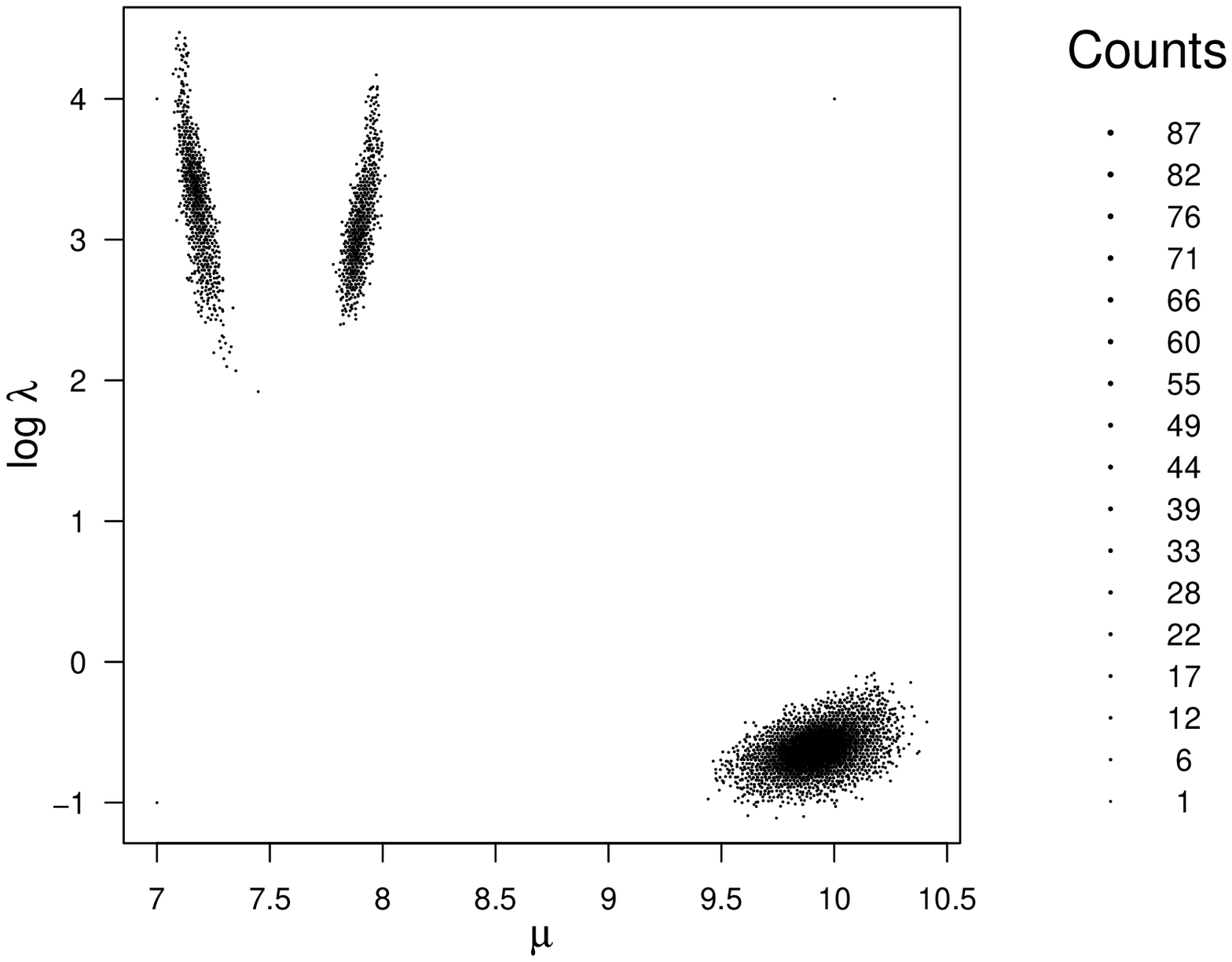}
  \includegraphics[width=0.3\textwidth]{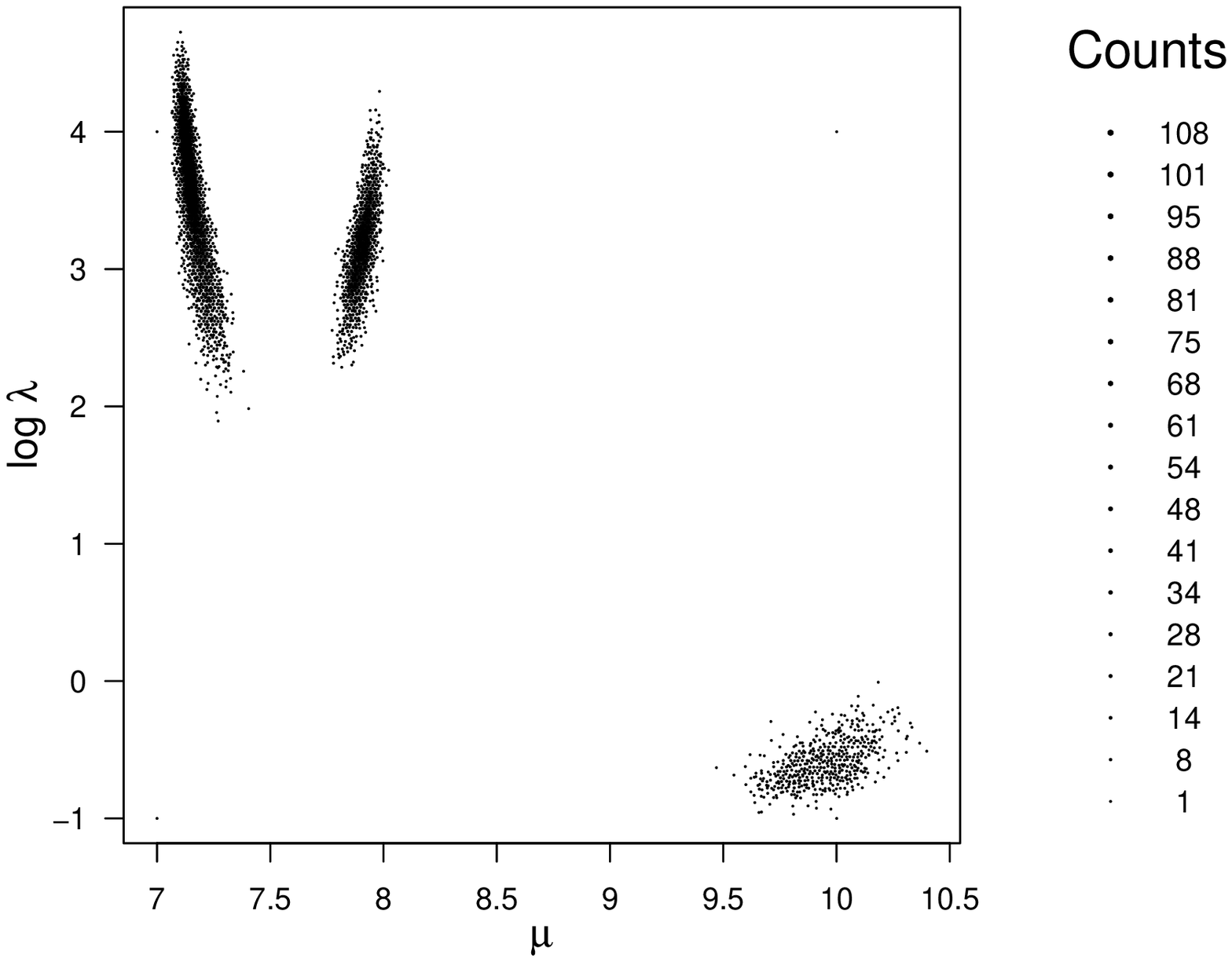}
\end{center}
 \caption{\label{traceplots3} Hexagon binning for
   $(\mu_k,\log\lambda_k)$, $k=1$, 2, 3,  for the free energy SMC
   sampler,  after  the final debiasing step, IBIS strategy.}
 \end{figure}

\begin{figure}[H]
\begin{center}
  \includegraphics[width=0.4\textwidth]{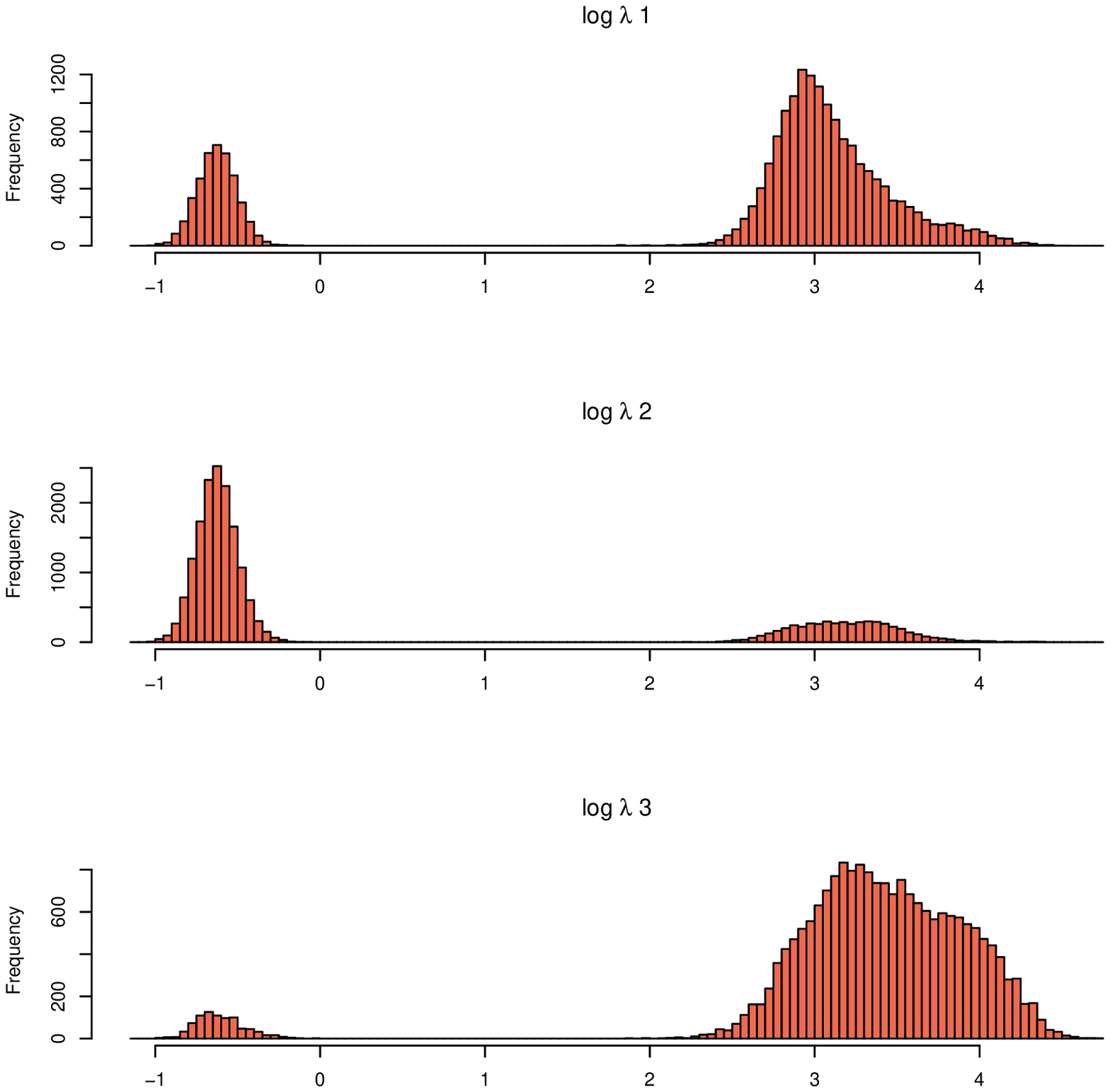}
  \includegraphics[width=0.4\textwidth]{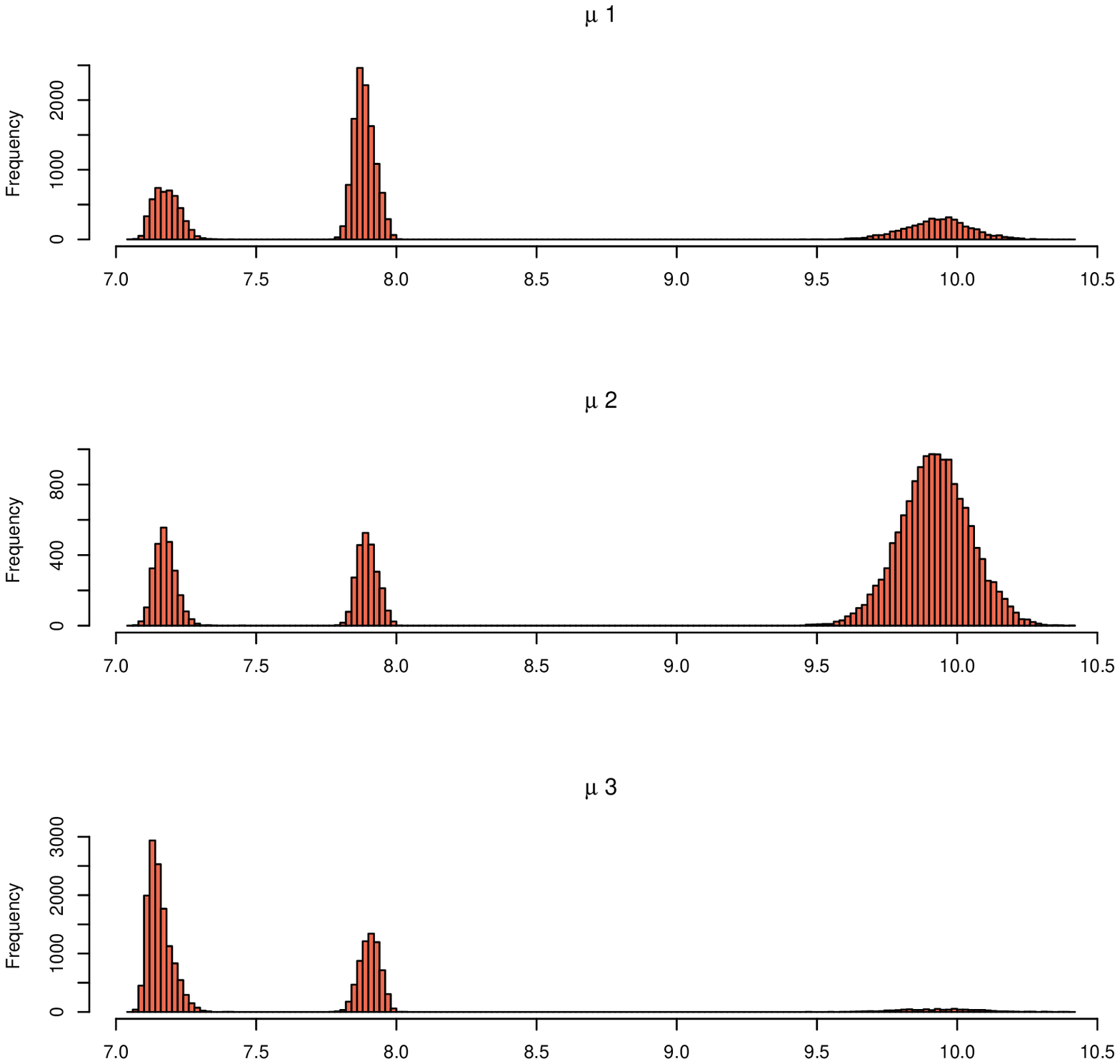}
\end{center}
 \caption{\label{histograms3}  Histograms of the components of the
   simulated particles obtained by free energy SMC
   sampler,  after  the final debiasing step, IBIS strategy.} 
 \end{figure}

 We then run a free energy SMC sampler, using the reaction coordinate
 $\xi$, 50 bins, and the ABP strategy for estimating the free
 energies. Figures \ref{traceplots2} and \ref{histograms2} represent
 the cloud of particles before the final unbiasing step, when the
 particles target the free energy biased density
 $\tilde{\pi}_T$. Figures \ref{traceplots3} and \ref{histograms3}
 represent the cloud of particles at the final step, when the target
 is the true posterior distribution. One sees that the output is not
 perfectly symmetric (at least after the final debiasing step), but at
 least the three equivalent modes have been recovered, and one can
 force equal proportions for the particles in each modal region, by
 simply randomly permuting the labels of each particles, if need be.

\begin{figure}[H]
\centering
\includegraphics[width=0.5\textwidth]{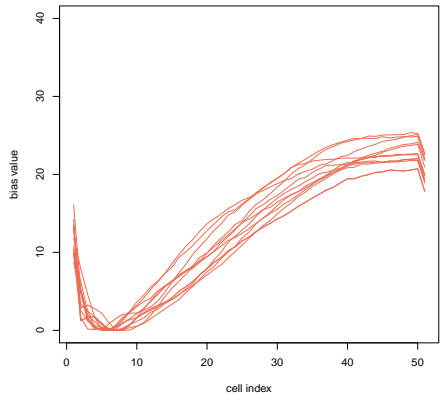}
\caption{\label{comparbias}Estimates of the final free energy $A_T$
  obtained from 10 runs of a free energy SMC sampler, versus cell indices}
\end{figure}

 To assess the stability of our results, we run the same sampler ten
 times, and plot the ten so-obtained estimates of the overall free
 energy $A_T$, which is used in the last debiasing step; see Figure
 \ref{comparbias}. Since a free energy function is defined only up to
 an additive function, we arbitrarily force the plotted functions to
 have the same minimum.

In short, one sees in this challenging example that (a) a nearly symmetric output
is obtained only if free energy biasing is implemented; and (b)
using free energy SMC, satisfactory results are obtained at a 
smaller cost than the adaptive MCMC sampler used in \cite{ChoLelSto}.

\subsection{Bivariate Gaussian mixtures}

\subsubsection{Prior, reaction coordinates}

We now consider a bivariate Gaussian mixture, $\psi(y;\xi)=N_{2}\left(\mu,Q^{-1}\right)$,
which is parametrised as follows:
\beqn
\xi_{k}=\left(\mu_{1,k},\mu_{2,k}, d_{1,k},d_{2,k},e_{k} \right),
\quad C_k = \left(
  \begin{array}{cc}
    d_{1,k}^{1/2} & 0 \\
   e_{k} & d_{2,k}^{1/2}
  \end{array}
\right), \quad Q_k = C_k C_k^T. 
\eeqn
This parametrisation is based on Bartlett decomposition: taking
$d_{1,k}\sim \mathrm{Gamma}(\alpha/2,\beta)$, 
$d_{2,k}\sim \mathrm{Gamma}((\alpha-1)/2,\beta)$, 
$e_k|\beta\sim N(0,1/\beta)$ leads to a Wishart prior for 
$Q_k$, $Q_k\sim\mathrm{Wishart}_2(\alpha,\beta I_2)$. This
parametrisation is also convenient in terms of implementing
the automatically tuned random walk Hastings-Metropolis strategy
discussed in Section \ref{sec:adaptivess-smc}. 

To complete the specification of the prior, we assume that \[
\mu_{k}=(\mu_{1,k},\mu_{2,k})'\sim N_{2}\left(M,S^{-1}\right),\quad\]
 that $\alpha=2$, and that
$\beta\sim \mathrm{Gamma}(g,h)$. Of course, this prior is meant to  generalise  the
prior used in the previous section in a simple way. In particular, 
the hyper-parameter $\beta$ should
play the same role as in the univariate Gaussian case, and we use it
as our reaction coordinate.

\subsubsection{Numerical results}

\begin{figure}[H]
\centering
\includegraphics[width=0.4\textwidth]{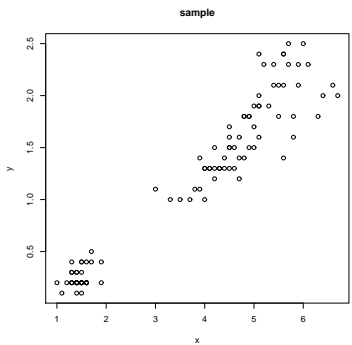}
\caption{\label{irissample} Iris sample}
\end{figure}

\begin{figure}[H]
\begin{center}
  \includegraphics[width=0.35\textwidth]{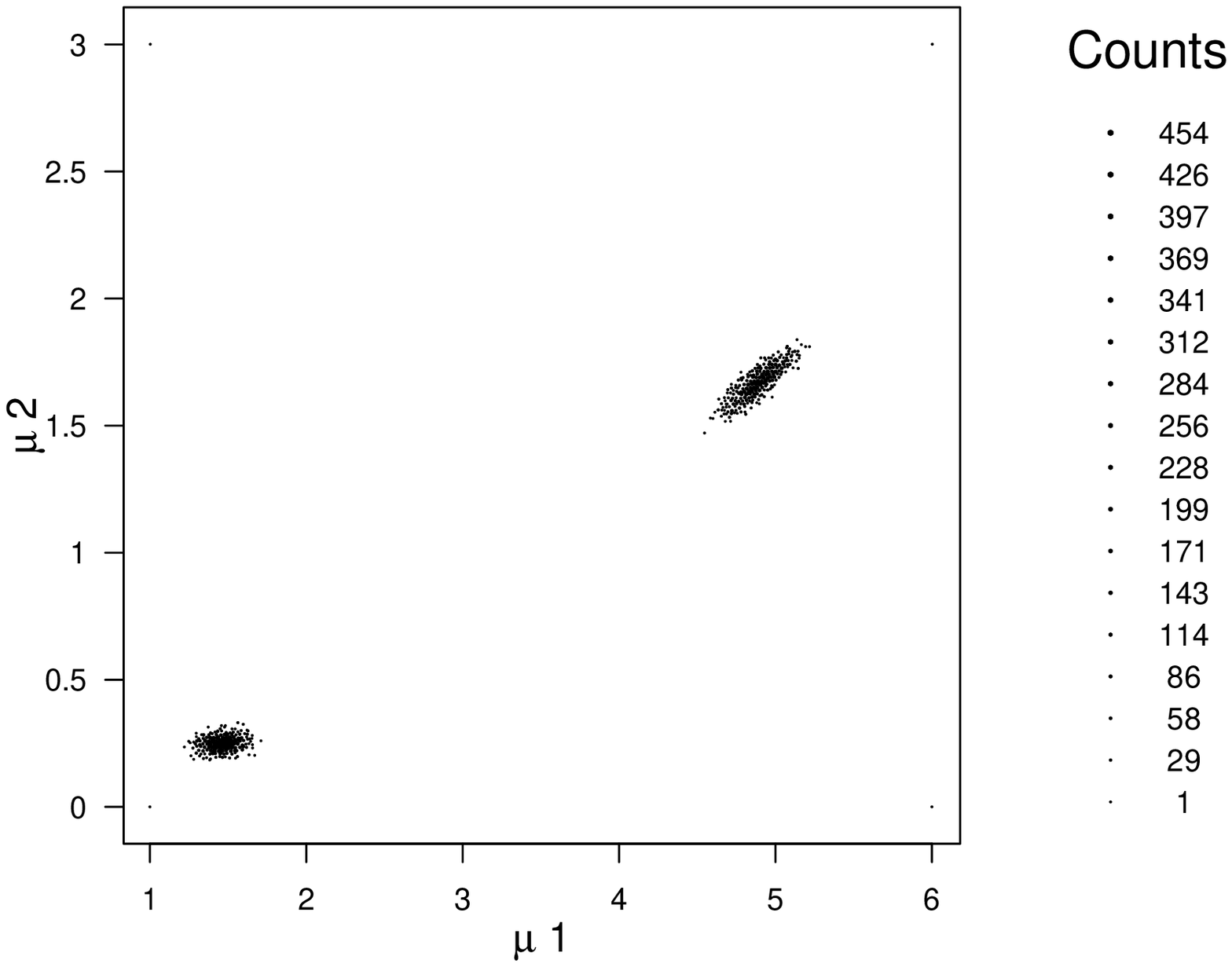}
  \includegraphics[width=0.35\textwidth]{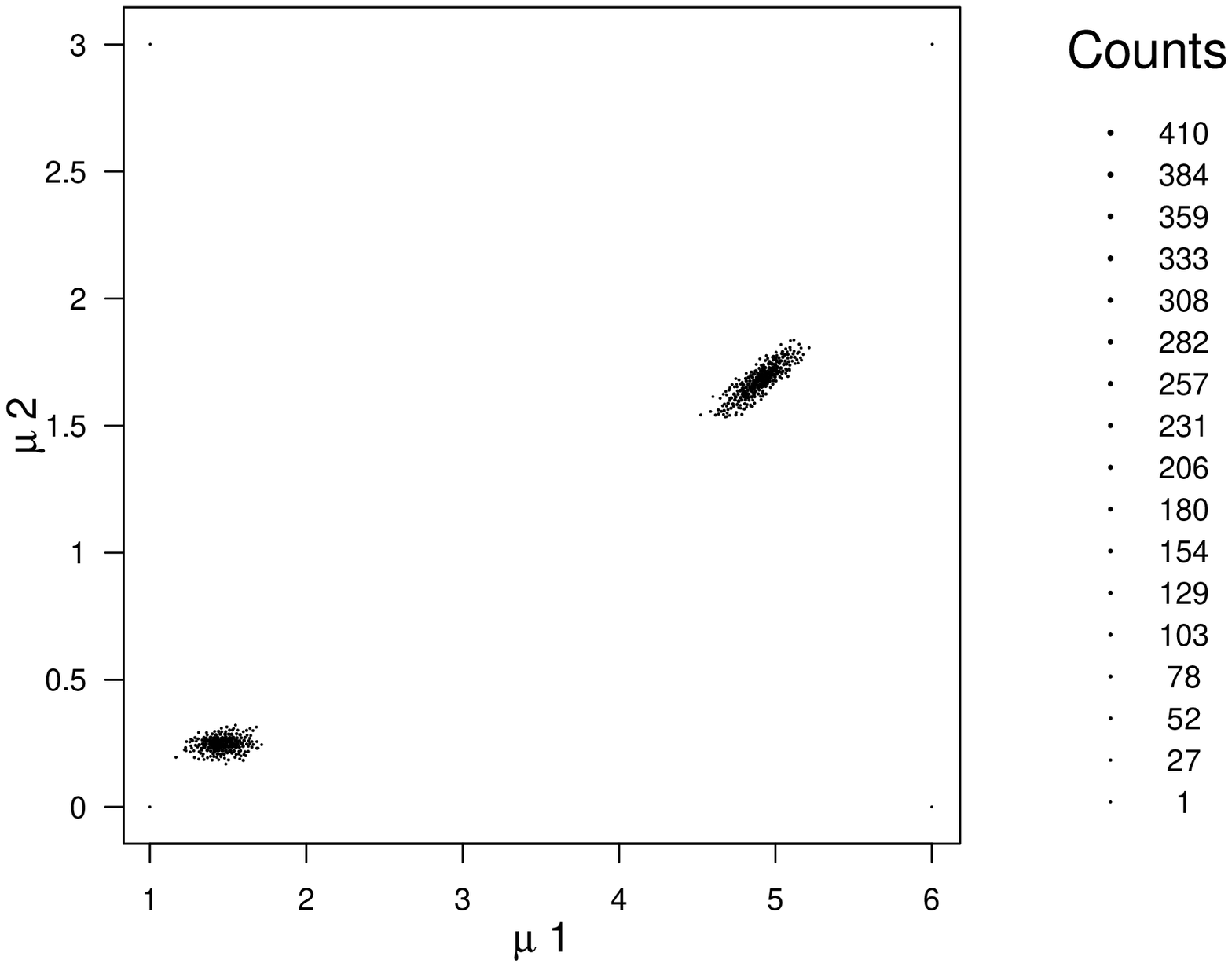}
\end{center}
 \caption{\label{multivartraceplots3} Hexagon binning for
   $\mu_k=(\mu_{k,1},\mu_{k,2})$, $k=1$, bivariate Gaussian example }
 \end{figure}

\begin{figure}[H]
\centering
\includegraphics[width=0.5\textwidth]{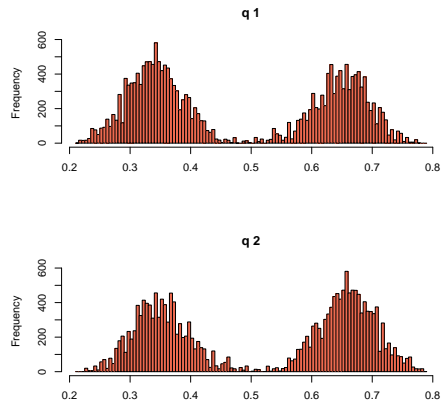}
\caption{\label{multivarhistweight3} Weighted histograms of
  $q_k=\omega_k/(\omega_1+\omega_2)$, for $k=1,$ 2,  bivariate Gaussian example }
\end{figure}

We consider two out of the four measurements recorded in Fisher's Iris
dataset, petal length and petal width, see
e.g. \citet[][Chap. 6]{Fru:book}, and Figure \ref{irissample} for a
scatter-plot. We take $K=2$. As in the previous example, we run a
standard SMC sampler (with the same number of particles, and so on),
and observes that only one mode is recovered.  We then run a free
energy SMC sampler.  For the sake of space, we report only the
debiased output at the very final stage of the free energy SMC
sampler, that is the cloud of particles targeting the true posterior
distribution. Figure \ref{multivartraceplots3} represents the
bivariate vectors $\mu_k$, and Figure \ref{multivarhistweight3}
represent the component probabilities
$q_k=\omega_k/(\omega_1+\omega_2)$ for $k=1$, 2. Clearly, the output
is nearly symmetric.

One sees in this example that free energy SMC still works well for
bivariate Gaussian mixture model, despite the larger dimension of the
parameter space. In particular, the choice of the reaction coordinate
seems to work along the same lines, i.e. choosing an hyper-parameter
that determines the spread of the components.

\section{Conclusion} \label{sec:Conclusion}

In this paper, we introduced free energy SMC sampling, and observed in
one mixture example that it may be faster than free energy methods
based on adaptive MCMC, such as those considered in \cite{ChoLelSto}.
It would be far-fetched to reach general conclusions from this
preliminary study regarding the respective merits of free energy SMC
versus free energy MCMC, or, worse, SMC versus Adaptive MCMC. If
anything, the good results obtained in our examples validates, in the
mixture context, the idea of combining two recipes to overcome
multimodality, namely (a) free energy biasing, and (b) tracking
through SMC some sequence $(\pi_t)$ of increasing difficulty, which
terminates at $\pi_T=\pi$.  Whether such combination should work or
would be meaningful in other contexts is left for further research.

\section*{Acknowledgements}

N.~Chopin is supported by the 2007--2010 grant ANR-07-BLAN ``SP Bayes".
P.~Jacob is supported by a PhD Fellowship from the AXA Research Fund.
 The authors thank Peter Green for  insightful comments. 

\bibliographystyle{apalike}
\bibliography{complete}

\end{document}